\documentclass[%
 aip,
 pop,
 amsmath,amssymb,
 preprint,%
]{revtex4-1}

\usepackage{graphicx}
\usepackage{dcolumn}
\usepackage{bm}
\usepackage{cleveref}
\usepackage{subcaption}
\usepackage{tabularx}

\usepackage[utf8]{inputenc}
\usepackage[T1]{fontenc}
\usepackage{mathptmx}
\usepackage{etoolbox}

\usepackage{amsmath}
\usepackage{cleveref}
\usepackage{todonotes}
\usepackage{cancel}
\usepackage{ulem}
\usepackage{siunitx}

\newcommand{\aeq}{\begin{equation}}
\newcommand{\eeq}{\end{equation}}
\newcommand{\aeqn}{\begin{eqnarray}}
\newcommand{\eeqn}{\end{eqnarray}}
\newcommand{\aeqs}{\begin{equation*}}
\newcommand{\eeqs}{\end{equation*}}
\newcommand{\aeqns}{\begin{eqnarray*}}
\newcommand{\eeqns}{\end{eqnarray*}}
\DeclareOldFontCommand{\bf}{\normalfont\bfseries}{\mathbf}
\DeclareOldFontCommand{\rm}{\normalfont\rmfamily}{\textrm}
\newcommand{\vecb}[1]{{\textbf #1}}

\newcommand{\vpar}{{v_\parallel}}

\newcommand{\fder}[2]{\frac{d{#1}}{d{#2}}}

\def \d {\mathrm{d}}

\newcommand{\Ampere}{Amp\`ere}

\newcommand{\be}{\begin{equation}}
\newcommand{\ee}{\end{equation}}
\newcommand{\ba}{\begin{eqnarray}}
\newcommand{\ea}{\end{eqnarray}}
\newcommand{\bas}{\begin{eqnarray*}}
\newcommand{\eas}{\end{eqnarray*}}

\newcommand{\pard}[2]{\frac{\partial #1}{\partial #2}}

\makeatletter
\def\@email#1#2{%
 \endgroup
 \patchcmd{\titleblock@produce}
  {\frontmatter@RRAPformat}
  {\frontmatter@RRAPformat{\produce@RRAP{*#1\href{mailto:#2}{#2}}}\frontmatter@RRAPformat}
  {}{}
}%
\makeatother

\begin{document}
\pagestyle{empty}

\preprint{}----------------------------------------
\pagenumbering{gobble}
\title[]{Verification of the PICLS electromagnetic upgrade in mixed variables}
\author{A. Stier}
 \email{annika.stier@ipp.mpg.de}
 \affiliation{Max-Planck Institute for Plasma Physics.}
\author{A. Bottino}
\affiliation{Max-Planck Institute for Plasma Physics.}
\author{D. Coster}
\affiliation{Max-Planck Institute for Plasma Physics.}
\author{T. Hayward-Schneider}
\affiliation{Max-Planck Institute for Plasma Physics.}
\author{L. Villard}
\affiliation{\'Ecole Polytechnique F\'ed\'erale de Lausanne}
\author{F. Jenko}
\affiliation{Max-Planck Institute for Plasma Physics.}

\date{\today}

\begin{abstract}
The gyrokinetic particle-in-cell code PICLS is a full-f finite element tool to simulate turbulence in the tokamak scrape-off layer.
During the previous year, the capability of PICLS was extended to encompass electromagnetic effects. Successful tests using the method of manufactured solutions were conducted on the freshly added \Ampere's-law-solver, and shear Alfvén waves were simulated to verify the new electromagnetic time step. 
However, as a code based on the $p_{||}$-formulation of the gyrokinetic equations, PICLS is affected by the \Ampere-cancellation problem. In order to bring higher-beta simulations within reach of our computational capacity, we implemented the mixed-variable formulation with pullback-scheme in a similar fashion to, e.g., EUTERPE, ORB5, or XGC. Here, we present the successful verification of the different electromagnetic formulations of PICLS  by simulating shear-Alfvén waves  in a test setup designed to minimize kinetic effects. 
\end{abstract}
\maketitle


\section{Introduction}
\begin{figure}
    \centering
    \includegraphics[width=\linewidth]{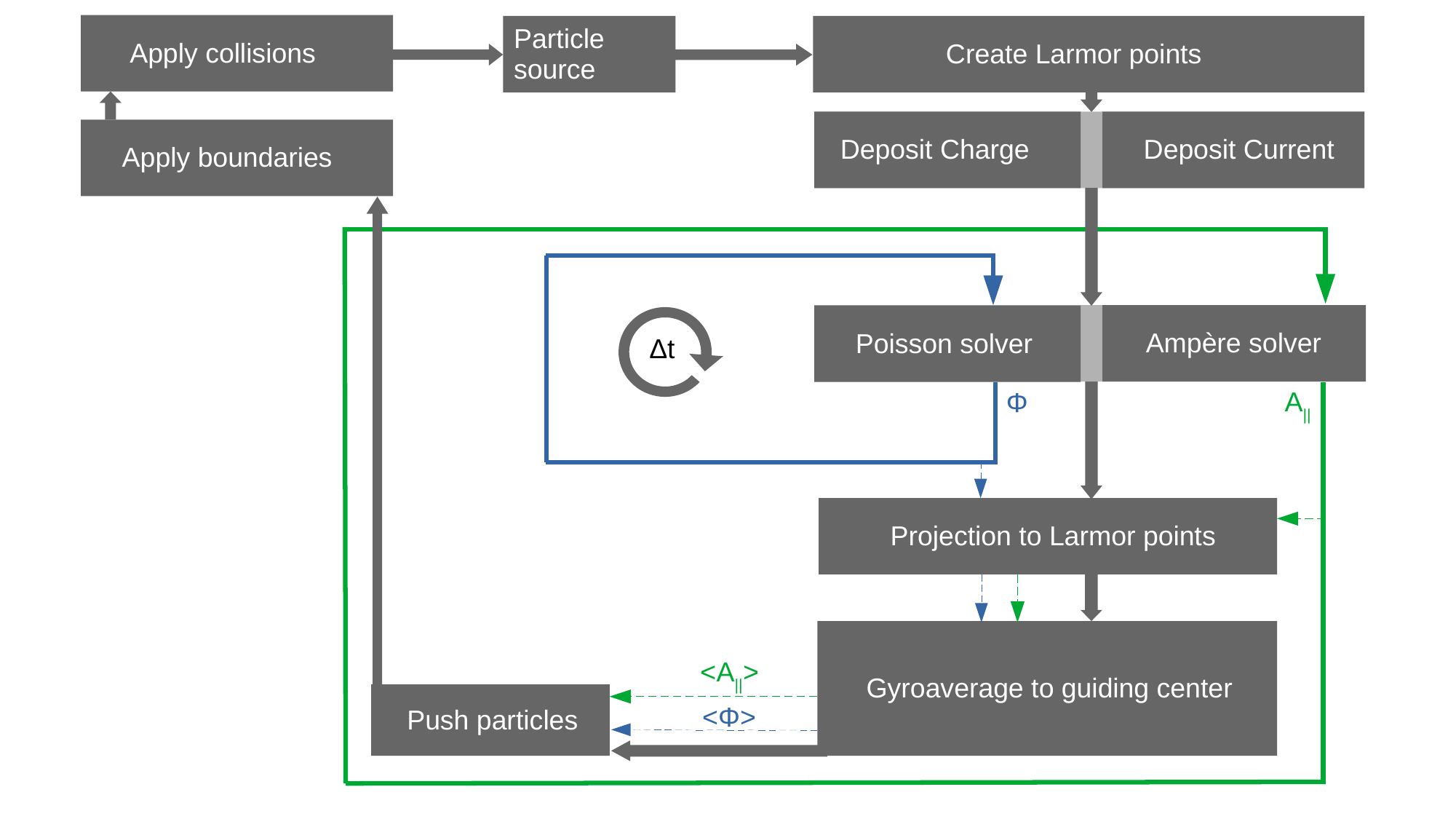}
    \caption{Overview over the time step of the electromagnetic PICLS code in $p_{\parallel}$-formulation without cancellation mitigation. Both potentials, $\Phi$ and $A_{\parallel}$, are updated each time step by the respective solvers and used in advancing the particle trajectories.}
    \label{fig:timestep}
\end{figure}

The full-f PICLS (Particle-In-Cell Logical Sheath) code is  a particle-in-cell code designed to solve the gyrokinetic equations in open field line geometry\cite{MatthiasPhD, bottino2024} in Cartesian coordinates.  PICLS discretizes the electric and magnetic potential fields using a b-spline-based finite element approach\cite{Stier2024}.  Being based on a full-f approach, PICLS is well suited for simulating plasmas with strong free energy sources (e.g. strong gradients) and non-local events (e.g. avalanches),  like those characterizing the tokamak edge and scrape-off layer. Each of the at least two species  present  in a PICLS simulation, one of which has to model the electrons while the others have to model ions, can be handled either gyrokinetically or driftkinetically.  The gyroaverage operator is discretized in real space in a parallelization-friendly way through so-called Larmor points\cite{Hatzky06}, similar to the approach used in the ORB5 code\cite{LantiCPC2020}.  Features of PICLS include optional particle sources, a Lenard-Bernstein collision operator\cite{Boesl2019}, optional application of a control variate and the namesake logical sheath boundaries.  In addition to the open field line geometry, PICLS allows for some closed field line configurations using polar coordinates.   Furthermore, the field solver of PICLS takes advantage of the presence of one periodic direction in the system. If the background magnetic field does not depend on the periodic direction (e.g., the toroidal angle in an axisymmetric device), the field solver can be highly parallelized by applying discrete Fourier transforms on the spline coefficients. Details can be found in our previous work\cite{Stier2024}.  
Relieving the electrostatic constraint in PICLS, we implemented an additional \Ampere-solver, modeled after the existing Poisson-solver, to calculate the parallel perturbed magnetic potential $A_{\parallel}$. As \cref{fig:timestep} shows, the electric and magnetic potentials are coupled only via the particles, so no modification to the Poisson-solver was necessary.
In sections 2 and 3, we present the \Ampere-solver and the electromagnetic equations of motion, both in the (Hamiltonian) $p_{\parallel}$-formulation, and with the mixed-variable pullback scheme\cite{mishchenko2014} implemented to mitigate  the \Ampere-cancellation problem. In implementing this scheme, we chose the same approach as other electromagnetic PIC codes like EUTERPE \cite{cole2016}, ORB5 \cite{mishchenko2019} or XGC \cite{hager2022}.\\  In the fourth  section, we will give a description of the initial particle loading since the method of sampling the marker attributes is fundamentally different compared to existing PIC delta-f codes like ORB5. In particular, we will focus on the Maxwellian loading needed to reproduce MHD-based results.  To verify the implementation, we conduct multiple tests, which are reported in section 5. We begin by verifying the implementation of the \Ampere-solver by means of the method of manufactured solutions, before showing the validity of the whole electromagnetic time step by reproducing shear-Alfvén-wave (SAW) physics in pinch geometry following the example of ORB5\cite{Biancalani2016}. Due to the gyrokinetic calculation constituting a main proposition of PICLS, we lastly add a demonstration of  the effect of the gyroaverage on ion-temperature gradient (ITG) instabilities in a cylindrical plasma. 

\section{Electromagnetic theory in $p_{\parallel}$-formulation}
\label{sec:ppar}
\subsection{Parallel \Ampere's law}
\label{sec:amperesolver}
 As already mentioned before \cite{bottino2024, Stier2024}, the physical model of PICLS is derived from the following gyrokinetic particle Lagrangian (written in CGS units):  

\begin{equation}   
        L=\sum_{s}\int \left(\left(\frac{q_s}{c}\vecb{A}+p_{\parallel}\vecb{b}\right)\cdot\Dot{\vecb{X}}+\frac{m_s c}{q_s}\mu\Dot{\Theta}-H_s \right)f_s \d W \d V + \int \frac{\tilde{E}^2-\tilde{B}_{\perp}^2}{8\pi} \d V 
        \label{eq:GKLagrangian}
\end{equation}
for particle species $s$ with charge $q_s$, mass $m_s$, parallel momentum $p_{\parallel}$, at gyrocenter position $\vecb{X}$  and gyro angle $\Theta$     in a phase space composed of the velocity space $W$ and physical space $V$. $\vecb{A}$ is the magnetic vector potential, $\vecb{b}$ the magnetic field unit vector, $\mu$ the magnetic moment, $f_s$ the distribution function  of species $s$,  $\Tilde{E}$ the perturbed electric field strength, $\Tilde{B}$ the magnetic field strength and $c$ the speed of light.
A general overview of the gyrokinetic theory and its application to plasma turbulence  has been covered in the literature \cite{ScottV2} The full derivation of \cref{eq:GKLagrangian} is likewise described in detail  elsewhere \cite{Tronko16}. The physics content of \cref{eq:GKLagrangian} depends on the choice of the Hamiltonian $H_s$. Here we use the following Hamiltonian,  in which electrostatic perturbations are assumed to have long perpendicular wavelengths as compared to the ion thermal Larmor radius \cite{Bott15} and only parallel perturbation in the magnetic potential are considered,
\begin{align}
    H_s =& H_{s,0} + H_{s,1} + H_{s,2},\\
    H_{s,0} =& \frac{p_{\parallel}^2}{2m_s} + \mu B,\\
    H_{s,1} =& q_s J_{s,0}\left( \Phi - A_\parallel \frac{p_\parallel}{m_sc}\right),\\
    H_{s,2} =& - \frac{m_s c^2}{2B^2}|\nabla_{\perp}\Phi|^2+\frac{q_s^2}{2m_sc^2}(J_{s,0}A_\parallel)^2,
\end{align}
where $J_{s,0}$ is the gyroaveraging operator and $\Phi$ the electrostatic potential.\\
Using the quasi-neutrality approximation, $E^2 \ll E_{E \times B}$ (assuming that the energy associated to the magnetic perturbation is much smaller than the energy associated to the $E\times B$ motion), and the so-called linearized polarization approximation ($f_s=f_{M,s}$ when multiplying  $H_{s,2}$), we obtain the following Lagrangian
\begin{align}
    \label{eq:simpleL}
    L =& \sum_{s} \int \left( \left( \frac{q_s}{c} \vecb{A} + p_{\parallel} \vecb{b} \right) 
    \cdot \Dot{\vecb{X} } + \frac{m_{s} c}{q_{s}} \mu \Dot{\Theta} -H_{s,0}-H_{s,1} \right) 
    f_{s} \d W \d V\nonumber\\
    +& \sum_{s} \int \left(\frac{m_{s} c^{2}}{2B^2} |\nabla_{\perp} \Phi|^{2} - \frac{q_s^2}{2m_s c^2}(J_{s,0}A_\parallel)^2\right)f_{M,s}  \d W \d V - \int \frac{|\nabla_\perp A_\parallel|^2}{8\pi} \d V,
\end{align}
in which now $\vecb{A}$ and $\vecb{B}=B\vecb{b}$ refer to the background magnetic field only.

The equations governing the evolution of the electrostatic and parallel magnetic potentials are constructed by setting the functional derivative of $L$ with respect to $\Phi$ and $A_\parallel$, respectively,  to zero in order to minimize the action integral (see Chapter 5.9 of ref.~\onlinecite{ScottV2}). 
The discretization of the gyrokinetic Poisson equation of PICLS has been discussed in detail in ref.~\onlinecite{Stier2024}. Here we focus on the parallel \Ampere's law, obtained by Taylor expanding the term  $(J_{s,0}A_\parallel)^2$ (see, e.g. ref.~\onlinecite{LantiCPC2020})
\begin{equation}
    \sum_s \frac{n_{0,s} q_s^2}{m_sc}A_{||} + \frac{c}{4\pi}\nabla_{\perp}^2 A_{||} - \sum_{s={\rm ions}}\frac{T_{0,s}n_{0,s}c}{4B^2}\nabla_{\perp}^2 A_{||} = \sum_s \int \frac{q_s p_{\parallel}}{m_s}J_{0,s}f_s dW
    \label{eq:ampere_full}
\end{equation}

Note that in this work the Finite Larmor Radius (FLR) correction corresponding to the third term of \cref{eq:ampere_full}  is neglected and cylindrical coordinates are used. i.e.~$\d V = r \d r \d \theta \d \varphi$ and
\begin{equation}
    \nabla_{\perp} = \nabla r \frac{\partial}{\partial r} + \nabla \theta \frac{\partial}{\partial \theta}.
\end{equation}

Therefore, the parallel \Ampere's law used in this work is
\begin{equation}
    \sum_s \frac{n_{0,s} q_s^2}{m_sc}A_{||} + \frac{c}{4\pi}\nabla_{\perp}^2 A_{||} = \sum_s \int \frac{q_s p_{\parallel}}{m_s}J_{0,s}f_s dW.
    \label{eq:ampere}
\end{equation}

The natural equation to solve using finite elements is the so-called weak form of \cref{eq:ampere}, which is built with multiplying \cref{eq:ampere} by a test function
\begin{equation}
    \tilde{\Lambda}_t(\vecb{x})=\Lambda_{t,j'}(r)\Lambda_{t,k'}(\theta)\Lambda_{t,l'}(\varphi),
\end{equation}

and integrating over the full volume. $\tilde{\Lambda}_t(\vecb{x})$ represents a tensor product of  piecewise polynomial 1D B-splines  in cylindrical coordinates $(r,\theta,\varphi)$. The weak form therefore is

\begin{equation}
    \int \left(\sum_s K_{1,s}(r)A_{\parallel}+K_2\nabla_{\perp}^2A_{\parallel}\right)\tilde{\Lambda}_t(\vecb{x}) r \d r \d\theta \d\varphi = \int \sum_s \int P_s(p_\parallel) J_{0,s}f_s dW \tilde{\Lambda}_t(\vecb{x}) r \d r \d\theta \d\varphi,
    \label{eq:ampere_weak}
\end{equation}
with 
\begin{equation}
    K_{1,s}(r) = \frac{n_{0,s}(r)q_s^2}{m_s c};~~K_2 = \frac{c}{4\pi}; ~~P_s(p_\parallel) = \frac{q_s p_{\parallel}}{m_s}.
    \label{eq:skinterm}    
\end{equation}

$A_{\parallel}$ in \cref{eq:ampere_weak} is expressed in terms of B-spline polynomials, of order $p$, with coefficients $a_{\parallel,jkl}(t)$
\begin{equation}
    A_{\parallel} = \sum_{l=0}^{n_{\varphi}-1}\sum_{j=0}^{n_{r}+p-1}\sum_{k=0}^{n_{\theta}-1} a_{\parallel,jkl}(t) \tilde{\Lambda}_w(\vecb{x}),
    \label{eq:apll_splinecoef}
\end{equation}
with
\begin{equation}
\tilde{\Lambda}_w(\vecb{x})=\Lambda_{w,j}(r)\Lambda_{w,k}(\theta)\Lambda_{w,l}(\varphi).
\end{equation}

The basis spline coefficients $a_{\parallel,jkl}$ constitute a discrete field on a grid  of $(n_r, n_{\theta}, n_{\varphi})$ points , which can be Fourier-transformed in $\varphi$, leading to the discrete Fourier transformed coefficients
\begin{equation} 
    a_{\parallel,jkl}(t)= \sum_{n=0}^{n_{\varphi}-1} a_{\parallel,jk}^{(n)}(t)\exp{\left(\frac{2\pi i}{n_{\varphi}}nl\right)}
\end{equation}

with $n$ being the toroidal mode number. Inserting this into \cref{eq:apll_splinecoef} leads to
\begin{align}
    \label{eq:5.35}
    A_\parallel =& \sum_{l=0}^{n_{\varphi}-1}\sum_{j}\sum_{k}\sum_{n=0}^{n_{\varphi}-1}a_{\parallel,jk}^{(n)}(t)\exp{\left(\frac{2\pi i}{n_{\varphi}}nl\right)} \Lambda_{j}(r)\Lambda_{k}(\theta)\Lambda_{l}(\varphi)
\end{align}

This expression is inserted into the weak formulation of the \Ampere's law, which is then integrated by parts. The elliptic structure of this equation implies that boundary conditions have to be applied in the non periodic directions. In the specific case considered in this paper, the radial direction $r$ is the only non periodic one and homogeneous Dirichlet boundary conditions are assumed on both sides of the radial domain. Note that, in general, PICLS allows for up to two nonperiodic coordinates in which  Dirichlet (zero and nonzero) boundary conditions can be applied.
The discrete weak \Ampere's law becomes
\begin{equation}
    \sum_{n=0}^{n_{\varphi}-1}B_{j'k'}^{(n)}\exp{\left(\frac{2\pi i}{n_{\varphi}}nl'\right)}=\sum_{n=0}^{n_{\varphi}-1}b_{j'k'}^{(n)}\exp{\left(\frac{2\pi i}{n_{\varphi}}nl'\right)}
\end{equation}
with
\begin{align}
    B_{j'k'}^{(n)} =& M^{(n)}\sum_{j}\sum_{k}a_{\parallel,j'k'}^{(n)} A_{j'k',jk}
\end{align}
having defined the {\it stiffness matrix} of elements $A_{j'k',jk}$, defined as
\begin{align}
    \label{eq:5.74}
    A_{j'k',jk} =& \int  \Bigl(\sum_s K_{1,s}(r,\theta)\nabla_{\perp}(\Lambda_{j'}(r)\Lambda_{k'}(\theta))\cdot\nabla_{\perp}(\Lambda_j(r)\Lambda_{k}(\theta))\\ 
    +& K_2 \Lambda_{j'}(r)\Lambda_{k'}(\theta)\Lambda_j(r)\Lambda_{k}(\theta)\Bigl) r \d r \d  \theta
\end{align}
corresponding to a set of $n_\varphi$ independent matrix equations, one for each Fourier mode
\begin{align}
\label{eq:finalsystem}
  \sum_{j}\sum_{k} A_{jk,j'k'}a_{\parallel,j'k'}^{(n)}=\frac{1}{M^{(n)}}b_{j'k'}^{(n)}
\end{align}
The coefficients $b_{j'k'}^{(n)}$ are calculated at every time step by taking the discrete Fourier transform of the current density vector
\begin{equation}
\label{eq:AMform_r}
    b_{j'k'l'} = \sum_{s=e,i} \int P_s(p_\parallel) J_{0,s}f_s\Lambda_{j'}(r)\Lambda_{k'}(\theta)\Lambda_{l'}(\varphi)  r \d r \d\theta \d\varphi.    
\end{equation}
The so-called {\it mass matrices}, $M^{(n)}$, contain all the dependence in $\varphi$ and can be precalculated at the beginning of the simulations. Their definition and derivation can be found in Appendix A of ref.~\onlinecite{Stier2024}.
The matrix elements $A_{j'k',jk}$ are sparse and independent of $\varphi$ with a banded block structure of $2p+1$ block bands. The right-hand side of the parallel \Ampere's law contains the gyroaveraged particle current density and is calculated during the PIC cycle's charge/current deposition step.
After that, the final system  of $n_{\varphi}$ matrix equations, described by \cref{eq:finalsystem}, can be solved with the help of existing linear algebra packages to obtain values for the Fourier coefficients $a_{\parallel,j'k'}^n$.  Contrary to the full 3D problem in real space, our set of matrices is trivial to parallelize in the code implementation. This leads to a faster code execution for a problem of fixed resolution.
The potential $A_\parallel(r,\theta,\varphi)$ can finally be calculated at any point in space by using \cref{eq:apll_splinecoef}.

\subsection{Equations of motion}
 In the particle pusher stage of the PICLS time step, the equations of motion (Euler-Lagrange equations) are applied to the marker guiding centers. These equations are derived from the particle Lagrangian by applying the variational principle on the action integral.
The full derivation can be found in refs.~\onlinecite{Tronko16, Tronko18}. At the lowest order \cite{Tronko16}, the equations of motion are
\begin{align}
    \dot{\vecb{X}} =& \pard{H}{p_\parallel} \frac{\vecb{B^{*}}}{B_{\parallel}^{*}} +\frac{c}{q_s B B_{\parallel}^{*}}\vecb{B}\times \nabla H, \label{eq:dxdt_picls} \\
    \dot{p}_{\parallel} =& -\frac{\vecb{B}^{*}}{B_{\parallel}^{*}} \cdot \nabla H,
\end{align}
with 
\begin{align}
    \vecb{B}^{*} =& \vecb{B} + \frac{c}{q_s}p_{\parallel}\nabla \times \vecb{b},\\
    \nabla H =& \mu \nabla B + q_s \nabla J_{s,0}\Psi,\\
    \Psi=&\Phi - A_\parallel \frac{p_\parallel}{m_sc}.\label{eq:motion}
\end{align}
This set of equations is used in this work to update the marker position during the {\it push marker} stage of the PICLS loop (see Fig.~\ref{fig:timestep}).

\section{Electromagnetic theory in mixed variables}
\label{sec:MV}
\subsection{Parallel \Ampere's law}

\Ampere's law in $p_\parallel$ formulation, \cref{eq:ampere}, contains on the left-hand-side (LHS), a term proportional to $A_\parallel$, the so-called {\it skin term} which is not present in gyrokinetic theories using the standard gyro-center $v_\parallel$ as parallel velocity coordinate. In most physically relevant cases, the skin term is orders of magnitude larger than the other. The meaning of this term can be easily understood by formally replacing $p_\parallel$ with $p_\parallel=m_sv_\parallel + A_\parallel q_s/c $ in the right-hand-side (RHS) of \cref{eq:ampere}. Therefore, the skin term is canceled exactly by the term proportional to $A_\parallel q_s/c$ in the RHS. In PIC codes, the entire RHS is discretized with particles, while
the skin term is discretized on a grid. Therefore, the two terms do not cancel exactly, leading to a significant signal/noise ratio problem: a large number of markers is required to reproduce a physically irrelevant term, while the actual physics is contained exclusively in a much smaller term. Therefore, without a mitigation scheme for \Ampere's cancellation problem, a high number of markers is necessary to overcome the numerical error resulting from the mismatch between the grid-based discretization of the left-hand side of \cref{eq:ampere} and the Lagrangian marker discretization of the currents on the right-hand side. 

As seen from the skin term $K_1$ in \cref{eq:skinterm}, the problem becomes larger with increasing  electron density, and consequently electron $\beta$, and can be mitigated by
unphysically increasing the electron mass. 
To avoid prohibitive computational cost at higher densities or for a natural electron mass, the mixed-variable  formulation of the gyrokinetic  equations  was proposed in ref.~\onlinecite{mishchenko2014}. It  consists of splitting $A_{||}$ into a Hamiltonian component ($A_{||}^{(h)}$) and a symplectic part ($A_{||}^{(s)}$), named after the terms they appear in the   Lagrangian. The parallel velocity dynamics is now described by the {\it mixed parallel momentum} $p_m=m_sv_\parallel + A_\parallel^{(h)} q_s/c $.
The gyrokinetic mixed variable Lagrangian used in this work is 

\begin{align}
    \label{eq:simpleLmixed}
    L =& \sum_{s} \int \left( \left( \frac{q_s}{c} \vecb{A} + p_m\vecb{b} q_s J_{0,s} A_{\parallel}^{(s)} \vecb{b}\right) 
    \cdot \Dot{\vecb{X} } + \frac{m_{s} c}{e_{p}} \mu \Dot{\Theta} -H_{s,0}-H_{s,1} \right) 
    f_{s} \d W \d V\\
    +& \sum_{s} \int \left(\frac{m_{s} c^{2}}{2B^2} |\nabla_{\perp} \Phi|^{2} - \frac{q_s^2}{2m_sc^2}(J_{s,0}A_\parallel^{(h)})^2\right)f_{M,s}  \d W \d V - \int \frac{|\nabla_\perp A_\parallel|^2}{8\pi} \d V\nonumber,
\end{align}
with
\begin{align}
    H_s =& H_{s,0} + H_{s,1} + H_{s,2},\\
    H_{s,0} =& \frac{p_m^2}{2m} + \mu B,\\
    H_{s,1} =& q_s J_{s,0}\left( \Phi - A_\parallel^{(h)} \frac{v_\parallel}{c}\right),\\
    H_{s,2} =& - \frac{m_s c^2}{2B^2}|\nabla_{\perp}\Phi|^2+\frac{q_s^2}{2m_sc^2}(J_{s,0}A_\parallel^{(h)})^2,\\
    A_\parallel =& A_\parallel^{(h)}+A_\parallel^{(s)}.
\end{align}
Using the same approximations as in the $p_\parallel$  formulation , the mixed variable parallel \Ampere's law thus becomes 
\begin{equation}
    A_{\parallel}^{(h)} \sum_s \frac{q_s^2 n_{0,e}}{m_s c} + \frac{c}{4\pi}\nabla_{\perp}^2 A_{\parallel}^{(h)} = \sum_s \int \frac{q_s}{m_s} p_m J_{0,s} f_s dW - \frac{c}{4\pi}\nabla_{\perp}^2 A_{\parallel}^{(s)}.
    \label{eq:ampereslaw_mv}
\end{equation}
The skin term depending on $A_{\parallel}^{(s)}$ on the right-hand side is now discretized on the grid and does not contribute to the cancellation problem.  Therefore, a cancellation problem is still present in this equation, but it can be kept small by ensuring $A_{\parallel}^{(h)}\ll A_{\parallel}^{(s)}$. Thus, the comparably small $A_{||}^{(h)}$ can be obtained with significantly lower marker counts as compared to the purely Hamiltonian case.

\subsection{Equations of motion}
The equations of motion in mixed variables depend on both $A_\parallel^{(h)}$ and  $A_\parallel^{(s)}$,
\begin{align}
    \dot{\vecb{X}} &=& \pard{H}{p_m} \frac{\vecb{B^{*}}}{B_{\parallel}^{*}} + \frac{c}{q_s B B_{\parallel}^{*}}\vecb{B}\times \nabla H - \frac{p_m}{cmB_{\parallel}^{*}} \vecb{b}\times\nabla A_\parallel^{(s)}\label{eq:dxdt_picls_mixed},\\
    \dot{p}_m &=& -\frac{\vecb{B}}{B_{\parallel}^*}\cdot \nabla \vecb{H}- q_s\frac{\partial A_{\parallel}^{(s)}}{\partial t} - \mu \frac{\vecb{b}\times \nabla \vecb{B}}{B_{\parallel}^*}\cdot \nabla A_{\parallel}^{(s)}
    \label{eq:dvdt_picls_mv},\\
    \vecb{B}^{*} &=& \vecb{B} + \frac{c}{q_s}p_{\parallel}\nabla \times \vecb{b} +  \nabla J_{0,s} A_\parallel^{(s)}\times \vecb{b}.
\end{align}
The parallel acceleration equation acquires a  dependence in $\partial A_{\parallel}^{(s)}/\partial t$ which will not allow for an explicit solver  to solve  the Vlasov-Maxwell system.
However, there is still a degree of freedom in the system, corresponding to the choice of the evolution equation for $A_\parallel^{(s)}$. A natural choice is
\begin{equation}
    \frac{\partial A_{\parallel}^{(s)}}{\partial t} = 0
\end{equation}
as described in ref.~\onlinecite{mishchenko2014}. The ideal MHD Ohm's law, $E_{\parallel} \approx 0$,  can also be used as discussed in Ref.~\onlinecite{mishchenko2014_2}. This choice is beneficial when MHD modes are simulated since it also relaxes restrictions on the time step, but it is not used in this work. 

\subsection{Pullback transformation}

The mixed variable splitting $A_\parallel=A_\parallel^{(h)}+A_\parallel^{(s)}$ does not necessarily imply $A^{(h)}\ll A^{(s)}$.

To ensure that, a pullback transformation is implemented that returns from the splitting of $A_{||}$ at the end of each time step back to a symplectic $v_{||}$-formulation by resetting $A_{||}^{(h,new)} = 0$ and storing the entire value of $A_{||}=A_{||}^{(s,old)}+ A_{||}^{(h,old)}$ in $A_{||}^{(s,new)}$. To be consistent with that, the pullback transformation needs to be applied to the particle velocities as well according to
\begin{equation}
    v_{\parallel,p}^{(new)} = v_{\parallel,p}^{(old)} - \frac{q_p}{m_pc} A_{\parallel,p}^{(h,old)}
\end{equation}

\begin{figure}
    \centering
    \includegraphics[width =\linewidth]{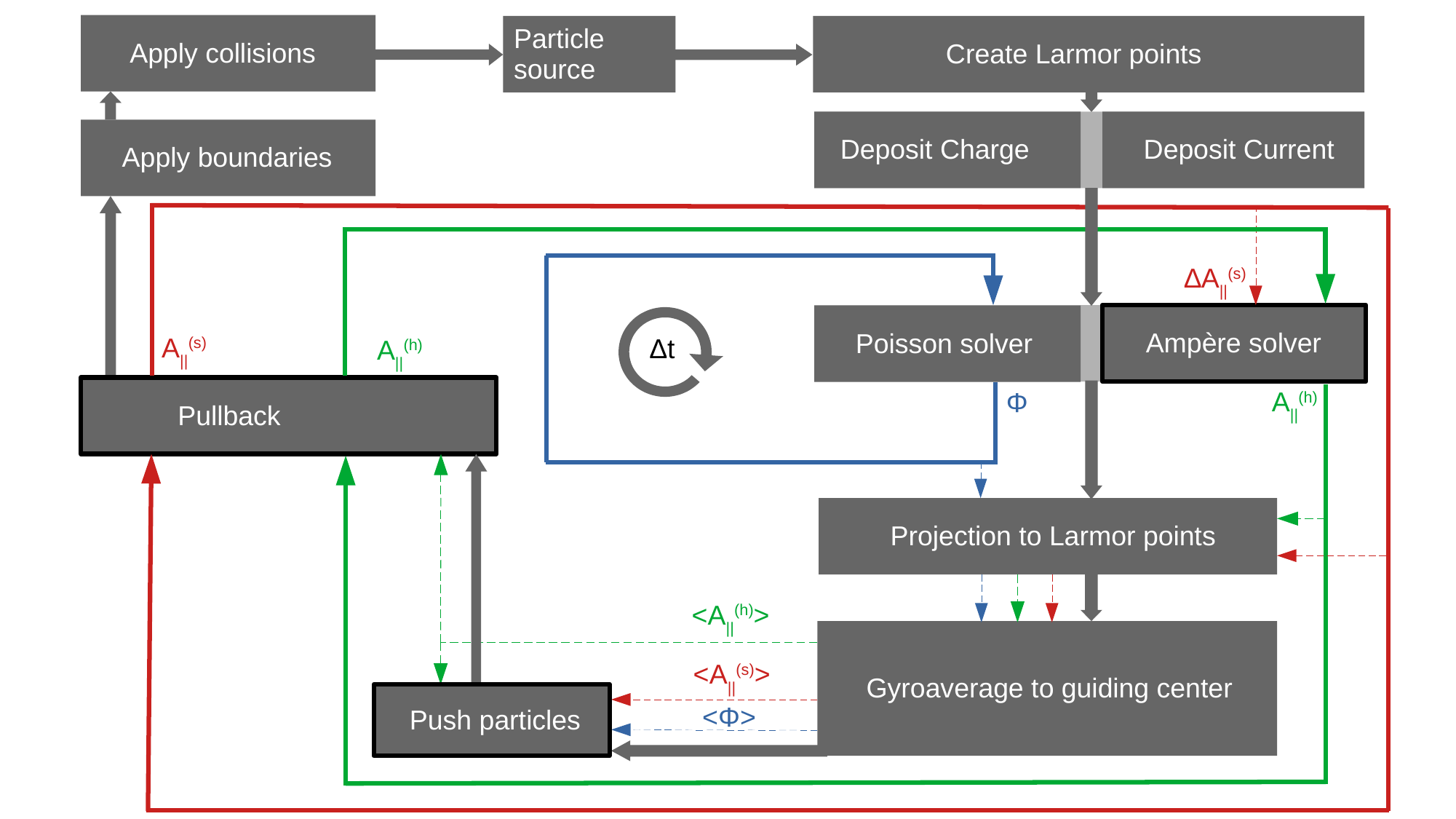}
    \caption{Overview over the time step of the electromagnetic PICLS code after implementing the mixed variable approach to mitigate \Ampere's cancellation problem. Compared to \cref{fig:timestep}, components that needed to be added or changed are marked with a black border. The Hamiltonian magnetic parallel vector potential $A_{\parallel}^{(h)}$ gets updated each time step by the \Ampere-solver and again in the pullback-transformation after being used for the pullback on the particle velocities. $A_{\parallel}^{(s)}$, on the other hand, is modified in the pullback only and is applied on the right-hand side of the \Ampere-solver.}
    \label{fig:timestep_mv}
\end{figure}

Note that this pullback numerical scheme is similar to the so-called {\it nonlinear pullback} of ref.~\onlinecite{mishchenko19}.
The complete PICLS electromagnetic time step, including the pullback-mitigation scheme, is displayed in \cref{fig:timestep_mv}.

\section{Initialization of the markers}
\label{sec:loading}
An important difference between PICLS and other existing PIC codes like ORB5 is that here, the entire gyrocenter distribution function $f(\vecb{X},v_\parallel,\mu)$
is represented by discrete markers. In this section, $v_\parallel$ indicates the generic variable used to describe the velocity along the field lines (e.g., $p_\parallel$). In the full-f representation, where the whole distribution will be simulated, the particle
distribution function thus can be expressed as
\aeq\label{eq:discretef}
f(\vecb{X} ,\vpar,\mu,t)=\frac{N_{\rm phys}}{N}\sum_{n=1}^{N}w_n(t) \delta(\vecb{X} -\vecb{X} _n(t))\delta(\vpar-\vpar_n(t))\delta(\mu-\mu_n),
\eeq
with $N$ the number of markers, $w_n$ the marker weights, $\vecb{X} _n$ their
position, $\vpar_n$ their parallel velocity and $\mu_n$ their (constant)
magnetic moment. The initial number of physical
particles is $N_\textrm{phys}=\int n_0(\vecb{X} )\d \vecb{X} =\bar{n}V$, having
defined the volume-averaged density
\aeq
\bar{n}=\frac{1}{V}\int n_0(\vecb{X} )\d \vecb{X} ,
\eeq
with $V$ being the space volume and $n_0$ the usual particle density.
In the absence of collisions, the weights for the full-f case are constant and thus do not change over time,
\aeq
\fder{}{t}w_n=0. 
\eeq
More details about the exact definition of the marker weights for PICLS and ORB5 can be found in ref.~\onlinecite{Bottino2022}.
Integrating \cref{eq:discretef} on a phase-space volume $\Omega_p$, centered
around a single marker which does not cross any other marker volume, we get
\aeq
f(\vecb{X_p},\vpar_p,\mu_p,t)\Omega_p=w_p(t).
\eeq
Therefore, the marker weight assumes the physical meaning of the number of
physical particles in the phase-space volume $\Omega_p$.
On the other hand, the volume $\Omega_p$ can also be defined as
\aeq\label{eq:omegap}
\Omega_p = \frac{\d N} {\d \vecb{X}  \d \vecb{v}},
\eeq
where $\d N$ is the number of markers contained in the infinitesimal volume $\d \vecb{X}  \d \vecb{v}$.

\begin{figure}
\centering
\includegraphics[width=\textwidth]{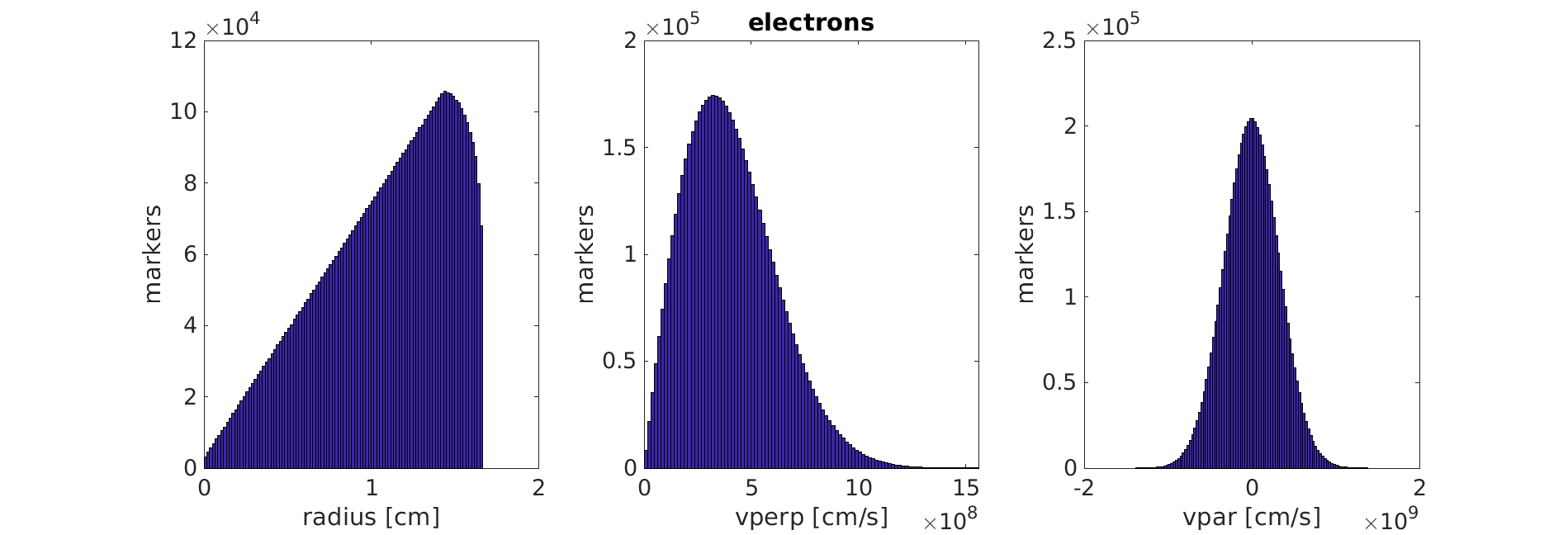}
\caption{Example of Maxwellian loading.
  Histograms in velocity and radius for a subset of 6 million test particles, loaded assuming
  a Maxwellian probability distribution function with constant $T$ and $n$
  profiles, showing that the loading is $\propto r$ in real space and $\propto
  v_\perp$ in velocity.}
\label{fig:maxwellian_loading}
\end{figure}

In the 3D version of PICLS, markers are loaded at $t=0$ in phase-space using importance
sampling, i.e.~according to  a probability distribution function $g$,
constructed in the following way. Given the infinitesimal phase-space volume $\d \vecb{X}  \d \vecb{v}$,
the number of physical particles in that volume is given by 
\aeq
\d N_{\rm phys} = f(\vecb{X} , \vecb{v}) \d \vecb{X}  \d \vecb{v},
\eeq
where $f(\vecb{X} , \vecb{v})$ is a generic  distribution
function. Although any physical equilibrium distribution function is an acceptable
initial distribution function for a full-f code, in most cases, $f(\vecb{X}, \vecb{v})$ is chosen to
be a Maxwellian distribution. Moreover, the choice of a Maxwellian is a requirement when comparing which MHD results, as is the case in this paper.
The number of markers contained in the same infinitesimal phase-space volume is given by
\aeq\label{eq:withg}
\d N = g(\vecb{X} , \vecb{v}) \d \vecb{X}  \d \vecb{v},
\eeq
where $g(\vecb{X} , \vecb{v})$ is a generic probability distribution. Comparing
\cref{eq:omegap} and \cref{eq:withg} implies $g(\vecb{X_p},\vecb{v_p})=1/\Omega_p$.
The distribution function $g$ can be obtained by imposing some requirements. The most natural choice is
\aeq
\frac{\d N_{\rm phys}}{\d N}=\frac{N_{\rm phys}}{N},
\eeq
corresponding to
\aeq\label{eq:withf}
\d N =   \frac{N}  {N_{\rm phys}} f(\vecb{X} , \vecb{v}) \d \vecb{X}  \d \vecb{v}.
\eeq
Comparing \cref{eq:withg} and \cref{eq:withf} we get
\aeq
g(\vecb{X} , \vecb{v}) = \frac{N}  {N_{\rm phys}} f(\vecb{X} , \vecb{v}),
\eeq
and for a single marker
\aeq
g(\vecb{X_p}, \vecb{v}_p) = \frac{N}  {N_{\rm phys}} f(\vecb{X_p}, \vecb{v}_p) 
= \frac{1}{\Omega_p}.
\eeq
As mentioned before, in all the cases considered in this paper, $f$ is assumed to be a Maxwellian, $f_M$,
and cylindrical coordinates $\vecb{X} =(r,\theta,z)$, or the equivalent $(r,\theta,\varphi)$ with $\varphi=2\pi z/L_z$, are used in real space.
Therefore, \cref{eq:withf} implies
\aeq\label{eq:loading}
\d N =   \frac{N}  {N_{\rm phys}} f_M(r, v_\parallel, v_\perp) (r \d r \d
\theta \d z) v_\perp (\d v_\perp \d v_\parallel \d \Theta),
\eeq
with
\aeq
f_M(r, v_\parallel, v_\perp)=n_0(r)K(r)\exp\left(\frac{m}{2}\frac{v_\parallel^2+v_\perp^2}{T_0(r)}\right),
\eeq
with $\Theta$ being the gyro-angle and $K(r)$ a  normalization factor. Moreover, \cref{eq:loading} shows that the marker loading follows a probability distribution
function $\propto rn_0$ in real space and $\propto v_\perp f_M$ in velocity space. 
An example of Maxwellian loading is illustrated in the histograms of \cref{fig:maxwellian_loading}, where
the marker distribution function has been reconstructed after the initial loading by binning a set of test particles (6M) in radius, $v_\perp$ 
and $v_\parallel$.\\

\section{Verification}
\subsection{Method of manufactured solutions}
\subsubsection{Method}
\begin{figure}
    \centering
    \includegraphics[width=\textwidth]{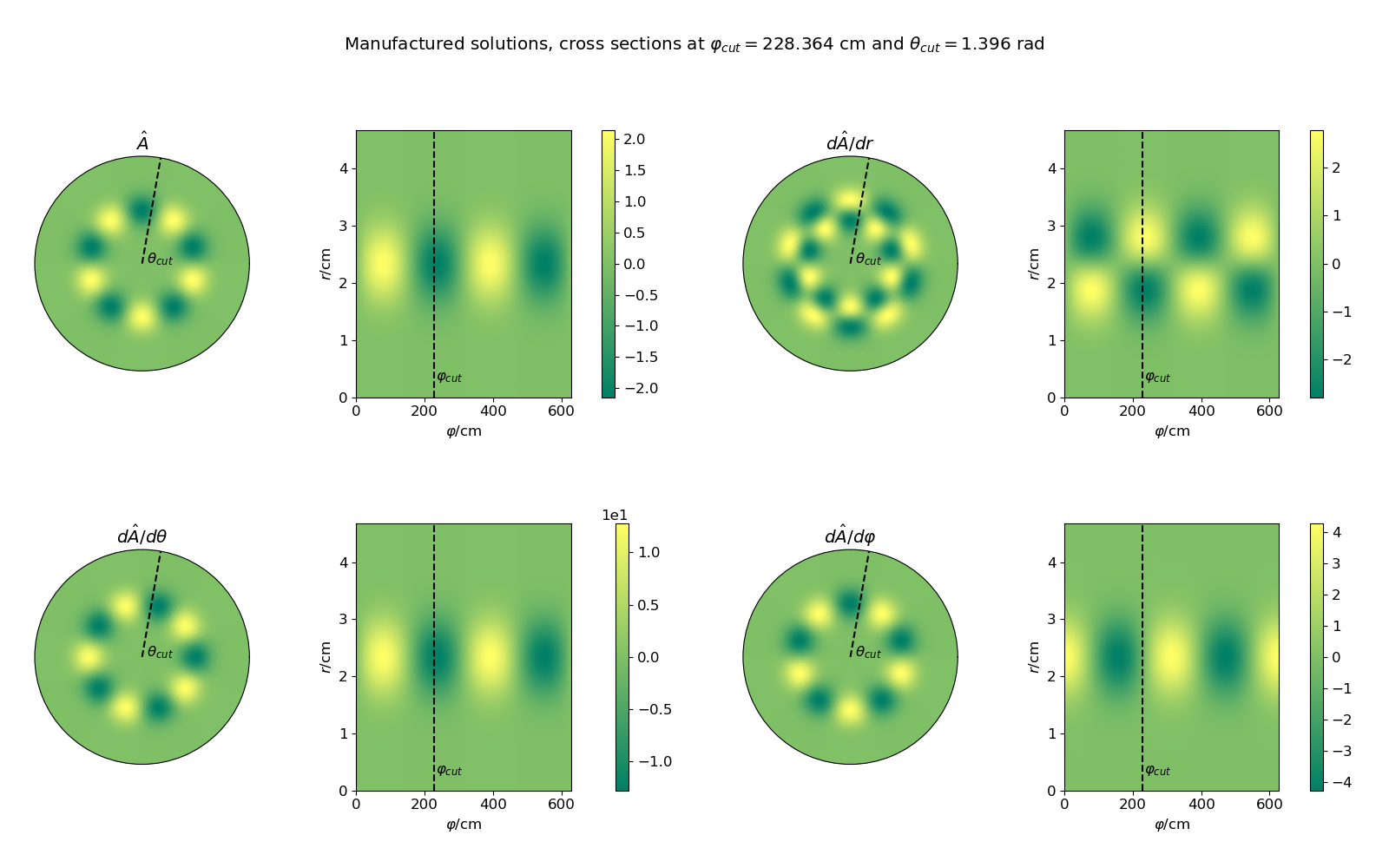}
    \caption{The manufactured solution $\hat{A}_{\parallel}$ \cref{eq:f1} and its gradients in $r$, $\theta$ and $\varphi$ with parameters $n_i=2$, $m_i=5$, $a=1$, $r_0=0.5r_{max}$, $\sigma_r=0.1r_{max}$, $\alpha=0$ and $\beta=-\exp{(-\frac{1}{2}(\frac{-r0}{\sigma_r})^2)}$ with $r_{max}$ the radial extension of the cylindrical domain as used for the MMS-test of the newly implemented $A_{\parallel}$-solver on a grid of (98,90,55) points.}
    \label{fig:anaHeatmap}
\end{figure}

The method of manufactured solutions \cite{Roache2001,Oberkampf10} is used here to verify the field solver for $A_{\parallel}$ (in $p_{\parallel}$-formulation) and $A_{\parallel}^{(h)}$ (in mixed variable formulation). It consists of using an analytical expression $\hat{A}_{\parallel}$ as a target \textit{manufactured solution}, from which a solver right-hand side is analytically obtained. This right-hand side, pertaining to $\hat{A}_{\parallel}$, is handed to the solver routine in place of a numerically calculated current density. The resulting parallel vector potential $A_{\parallel}$ can then be compared to $\hat{A}_{\parallel}$ to quantify an error. Lastly, it is checked that the error decays with increasing spatial resolution of the solver in adherence with the expected power law dependency of $p-1$. This method has previously been employed in our earlier works \cite{Stier2024,bottino2024} to verify the Poisson solver for the electrostatic potential $\Phi$.\\
For the test of the $A_{\parallel}$-solver, we choose an analytical function
\begin{equation}
    \hat{A}_{\parallel}(r,\theta,\varphi) = \sin{(n_i \varphi)}\sin{(m_i \theta)}a(\exp{\left(-\frac{1}{2}\left( \frac{r-r_0}{\sigma_r}\right) ^2\right)} +\alpha r +\beta)
    \label{eq:f1}
\end{equation}
with parameters $n_i$, $m_i$, $a$, $r_0$, $\sigma_r$, $\alpha$ and $\beta$ and dependency on all three spatial dimensions $r$, $\theta$ and $\varphi$. This is of importance, as we not only want to make sure that $A_{\parallel}$ is solved for correctly but also want to include $\nabla A_{\parallel}$ in our test, as $\nabla A_{\parallel}$ governs the electromagnetic impact on the particle trajectories as part of $\nabla\Psi$,  
\begin{equation}
    \nabla A_{\parallel} = \frac{\d A_{\parallel}} {\d r}\nabla r + \frac{\d A_{\parallel}} {\d\theta}\nabla \theta + \frac{\d A_{\parallel}} {\d\varphi}\nabla \varphi,
\end{equation}
with 
\begin{align}
    \frac{\d A_{\parallel}} {\d r}=& \sum_{l=0}^{n_{\varphi}-1}\sum_{j=0}^{n_{r}+p-1}\sum_{k=0}^{n_{\theta}-1} a_{\parallel,jkl}(t) \frac{\d \Lambda_{w,j}(r)}{\d r}\Lambda_{w,k}(\theta)\Lambda_{w,l}(\varphi),\\
    \frac{\d A_{\parallel}} {\d \theta}=& \sum_{l=0}^{n_{\varphi}-1}\sum_{j=0}^{n_{r}+p-1}\sum_{k=0}^{n_{\theta}-1} a_{\parallel,jkl}(t) \Lambda_{w,j}(r) \frac{\d \Lambda_{w,k}(\theta)}{\d \theta}\Lambda_{w,l}(\varphi),\\
    \frac{\d A_{\parallel}} {\d \varphi}=& \sum_{l=0}^{n_{\varphi}-1}\sum_{j=0}^{n_{r}+p-1}\sum_{k=0}^{n_{\theta}-1} a_{\parallel,jkl}(t) \Lambda_{w,j}(r) \Lambda_{w,k}(\theta) \frac{\d \Lambda_{w,l}(\varphi)}{\d \varphi}.\\
\end{align}
The manufactured solutions for $\nabla_r A_{\parallel}$, $\nabla_{\theta} A_{\parallel}$ and $\nabla_{\varphi} A_{\parallel}$ are trivially
\begin{align}
    \frac{d\hat{A}_{\parallel}}{dr} =& \sin{(m_i \theta)}\sin{(n_i \varphi)} a \left(\alpha - \left(\frac{r - r_0}{\sigma_r^2}\exp{\left(-\frac{1}{2}\left(\frac{r - r_0}{\sigma_r}\right)^2\right)}\right)\right)\\
    \frac{d\hat{A}_{\parallel}}{d\theta} =& m_i \cos{(m_i \theta)} \sin{(n_i \varphi)} a \left(\exp{\left(-\frac{1}{2}\left(\frac{r - r0}{\sigma_r}\right)^2\right)} + \alpha r + \beta\right)\\
    \frac{d\hat{A}_{\parallel}}{d\varphi} =& n_i \sin{(m_i \theta)} \cos{(n_i \varphi)} a \left(\exp{\left(-\frac{1}{2}\left(\frac{r - r0}{\sigma_r}\right)^2\right)} + \alpha r + \beta \right)\\
\end{align}
A graphic representation of the manufactured solutions is depicted for convenience in \cref{fig:anaHeatmap}.
We obtain the manufactured right-hand side
\begin{equation}
    \hat{\rho} = \sum_{s=i,e}K_{1}(r)\hat{A}_{\parallel}-K_2 \nabla_{\perp}^2\hat{A}_{\parallel}
    \label{eq:anaRHS}
\end{equation}
from the parallel  \Ampere's law by inserting \cref{eq:f1} for $A_{\parallel}$ and using the usual \Ampere-coefficients previously introduced in \cref{eq:skinterm}.
To be used by the solver, this must be projected onto the spline basis $\tilde{\Lambda}$, brought into the weak form, and be Fourier-transformed in the periodic direction $\varphi$, ending up with $\hat{\rho}_{jk}^{(n)}$. Since we used the same coefficients $K_1$ and $K_2$ as in a normal run, the \Ampere- matrix can be built as usual:
\begin{align}
    \hat{A}_{j'k'jk}&=\int \Bigl(K_1 \tilde{\Lambda}_{jkj'k'}(r,\theta) + \\
    &K_2(\nabla r)^2 \frac{\partial}{\partial r} \tilde{\Lambda}_{jk}(r,\theta)\frac{\partial}{\partial r} \tilde{\Lambda}_{j'k'}(r,\theta) + \\
    &K_2(\nabla \theta)^2 \frac{\partial}{\partial \theta} \tilde{\Lambda}_{jk}(r,\theta)\frac{\partial}{\partial \theta} \tilde{\Lambda}_{j'k'}(r,\theta)\Bigl) r \d r \d\theta
    \label{eq:Amperemat}
\end{align}
The solver then solves the matrix problem
\begin{equation}
    \frac{\hat{\rho}_{jk}^{(n)}}{M^{(n)}} = \sum_{j}\sum_{k} a_{\parallel,jk}^{(n)}(t) \hat{A}_{jkj'k'}
\end{equation}
for the coefficients $a_{\parallel,jk}^{(n)}$ needed to form the solution $A_{\parallel}$ and its derivatives.
After obtaining the numerical solution $A_{\parallel}$, the $L_2$-error over the whole domain is given by
\begin{equation}
    L_2=\frac{\sqrt{\sum_{ijk}(\hat{A}_{\parallel}(r_i,\theta_j,\varphi_k)-{A_{\parallel}}(r_i,\theta_j,\varphi_k))^2}}{\sqrt{\sum_{ijk}\hat{A}_{\parallel}(r_i,\theta_j,\varphi_k)^2}}
\end{equation}
For our solver test setup, we use a screw pinch configuration for the equilibrium field, with a magnetic field strength of $B_0=2$ T on the axis and a constant $q$-profile of $q_0=2$. The radial $r$-dimension spans from $0$ to $4.68$ cm  and is  bound by homogenous Dirichlet boundary conditions, while the poloidal $\theta$-dimension and the $\varphi$-dimension are both periodic. The $\varphi$-dimension is the one the solver will perform a discrete Fourier transform on. 
All tests were conducted in sequential mode with neither the MPI nor OpenMP parallelization options available in PICLS.

\subsubsection{Results}
\begin{figure}
    \centering
    \includegraphics[width=\textwidth]{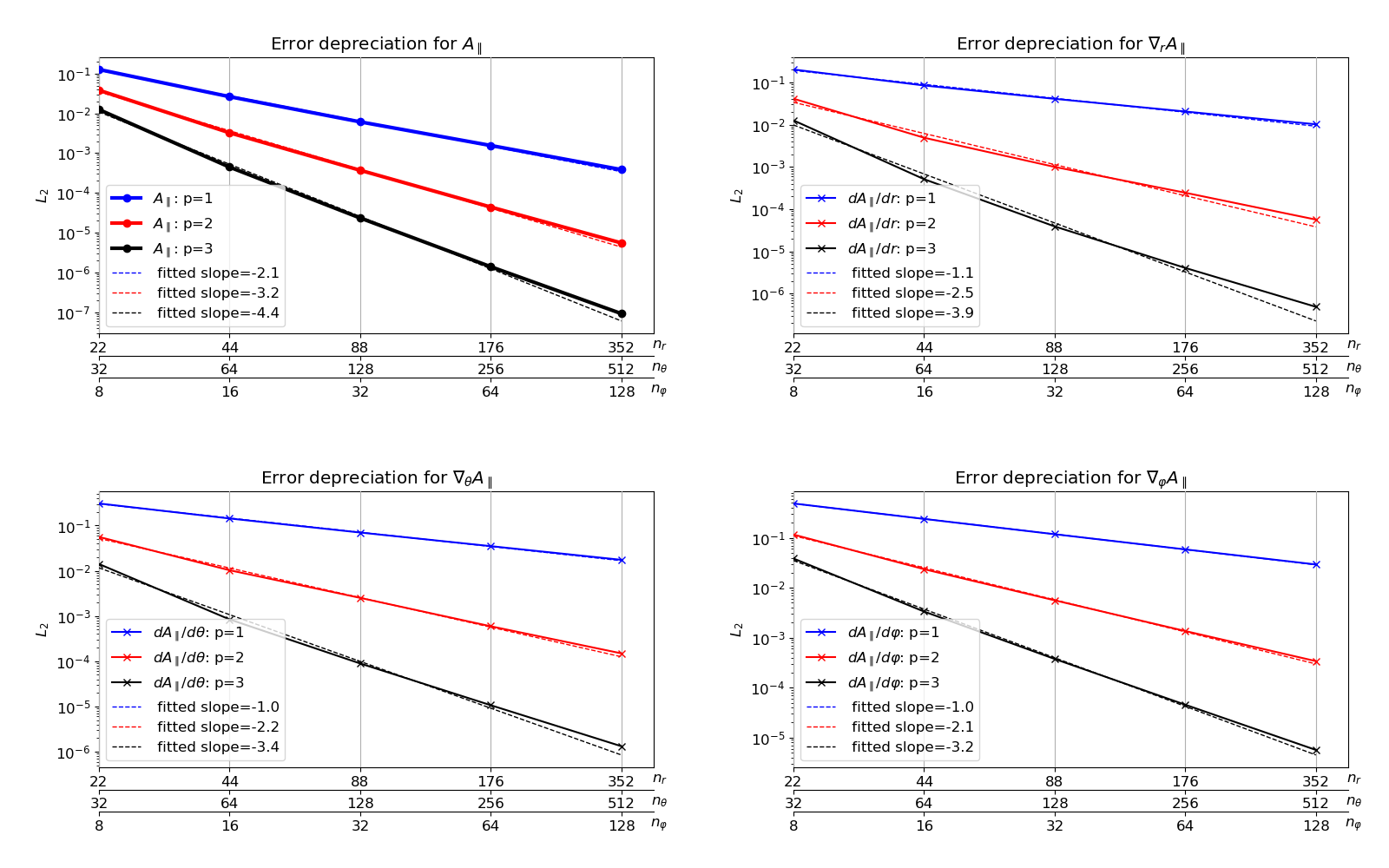}
    \caption{$L_2$-Error depreciation with increasing resolution for $A_{\parallel}$ and its spacial derivatives.}
    \label{fig:L2_mms}
\end{figure}

\begin{figure}
    \centering
    \includegraphics[width=\textwidth]{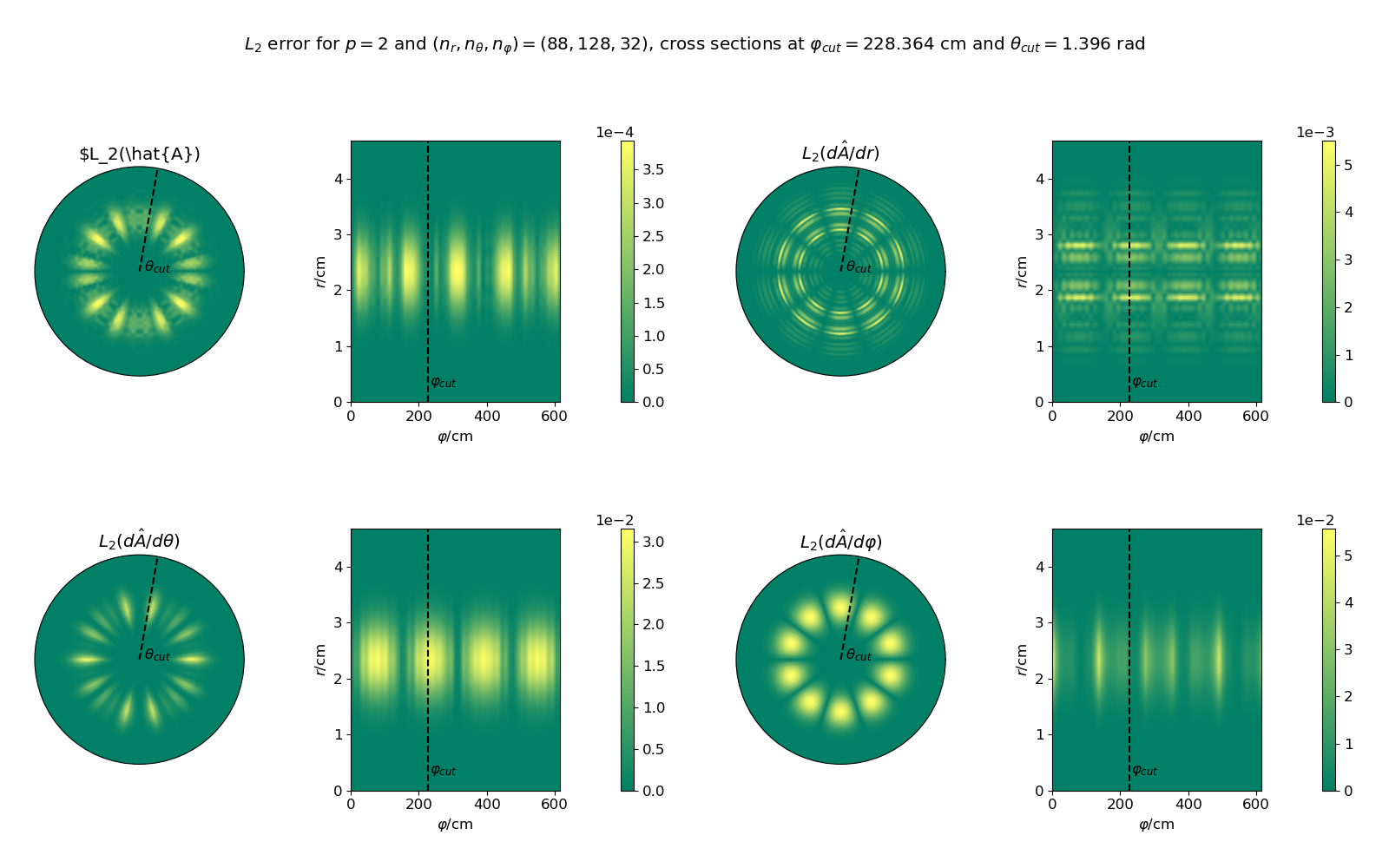}
    \caption{$L_2$ error distribution in the computational domain for $p=2$, $(n_r/n_{\theta}/n_{\varphi})=(88/128/32)$}
    \label{fig:errorHeatmap}
\end{figure}

In the double logarithmic plot \cref{fig:L2_mms}, the ideal slope of the $A_{\parallel}$ error curve is mathematically expected\cite{Li2010} to equal $-(p + 1)$ for spline degree $p$. In compliance with this, the error curves for the gradients of $A_{\parallel}$, computed with the derivatives of the basis splines and thus of degree $p-1$, ideally depreciate with a power law dependency of $-p$. \Cref{fig:L2_mms} shows that the deviation from this target does not surpass $1$ for any of the test series.
For an illustrative example of the spatial distribution of the error, we refer the reader to \cref{fig:errorHeatmap}.

\subsection{Shear-Alfvén-wave verification}
\label{sec:shearAlfven}
\subsubsection{Method}
An analytic expression of the MHD dispersion relation for a shear-Alfvén-wave (SAW) is given by
\begin{equation}
    \omega_{SAW} = v_{A}k_{\parallel} = \frac{B_0}{\sqrt{4\pi n_0 m_i}}\left(\frac{n}{R_0}+\frac{m}{R_0 q_0}\right)
    \label{eq:disprel_saw}
\end{equation}
for the Alfvén velocity $v_A$, parallel wave number $k_{\parallel}$, initial density $n_0$, ion mass $m_i$, toroidal and poloidal wave numbers $n$ and $m$, major radius $R_0$ and safety factor $q_0$.  We will compare the frequency of a SAW simulated with the newly implemented electromagnetic version of PICLS to this analytical prediction for different parameters $q$, $T_0$, and $n_0$ as a verification of our code. Note that an exact match can not be expected due to the MHD character of \cref{eq:disprel_saw}, which is not equivalent to the kinetic treatment of the plasma by PICLS.\\  Given that 
\begin{equation}
    \beta = \frac{8\pi n_0kT_0}{B_0^2}
\end{equation}
the scans in $n_0$ and $T_0$ are simultaneously scans in $\beta$ with increasing dominance of electromagnetic effects. Note that the skin term $K_1$ also increases with $n_0$, worsening the effects of \Ampere's cancellation problem.\\ The test series varying $n_0$ and $q_0$ closely follow an earlier investigation of ORB5\cite{Biancalani2016} recovering SAW dynamics. Supplementary, a scan in $T_0$ is performed  with the purpose of observing Landau-damping, which as a kinetic effect is not covered by the MHD model of \cref{eq:disprel_saw}. We choose the density for the $T_0$-scan such that the same $\beta$-values as is the $n_0$-scan are reached. In contrast to the density scan, we don't expect a shift in $\omega$. Instead, higher temperatures are expected to lead to a decay of the mode due to an increased level of Landau damping \cite{villard1995}.

\subsubsection{Setup}
This setup is chosen to be very similar to the one used in verifying ORB5 \cite{Biancalani2016}. 
Just like in this publication, we choose a cylindrical screw pinch of radius $r_{max}=1.67$ cm and a background magnetic field strength of $B_0 = 2.4$ T on the axis. The linear q-profile is subject to a parameter scan, the values of which are listed in \cref{tab:para}. The boundary conditions are periodic for the potential fields as well as the particles in the z-direction, and Dirichlet boundary conditions of $\Phi=A_{||}=0$ are applied at $r=r_{min}$ and $r=r_{max}$. The particle boundary condition on the outer radius $r_{max}$ resets any leaving particles' $r$-coordinate $r_p$ to a random value $r_{min} < r_p < r_{max}$. We use an initial density perturbation on the ions of shape
\begin{equation}
    \delta n = n_0(r)\left( 1 + \cos{(m \theta + n \varphi)}  \exp{-\frac{1}{2} \left(\frac{r - \frac{1}{2}r_{max}}{r_{max}  r_{width}}\right)^2}\right)
\end{equation}
with a poloidal mode number of $m=1$, a toroidal mode number of $n=0$, and a radial width parameter $r_{width} = 0.1$ while electrons are initialized with a homogeneous distribution. This provides the initial potential perturbation to start the SAW dynamic. It is worth mentioning that this initial perturbation also excites a sound wave, complicating the interpretation of the results. Therefore, the ion weight distribution is reset to a Maxwellian after the first call to the solver, thus suppressing the sound wave related to the initial ion pressure perturbation. The flat density profile $n_e=n_i$ is of varying magnitude and is detailed in \cref{tab:para}. The initial temperature profiles are equally flat and identical for both species. Values for $T_e=T_i$ are listed in \cref{tab:para} along with the other scan parameters. All tests are performed for three different mass ratios $m_i/m_e = 20$, $200$, and $2000$ to better compare our reference\cite{Biancalani2016}. While the ion species remains Deuterium for all test series, the electron mass is modified to meet these mass ratios. The time step for all simulations in the $n_0$- and $q_0$-scan is fixed at $\num{2.625e-8}$ s, which can be verified to be smaller than the required minimum time step listed in \cref{tab:para}. For the case of the temperature scan, a smaller time step of $\num{1e-9}$ s was prescribed to account for the high thermal velocities present in the highest $\beta$ case. The mixed variable scheme from \cref{sec:MV} has been used in all tests, and both species are treated drift kinetically with $\num{8e6}$ collisionless markers per species. To illustrate the detrimental effects of the cancellation problem and demonstrate their successful mitigation by the mixed variable formulation, we repeat the simulation with identical parameters in the $p_{||}$-version of the code. The fields are resolved with ($n_r$/$n_{\theta}$/$n_{\varphi}$)=(40/32/32) knots for splines of degree $p=2$. \\
 In order to reduce the statistical noise, an analytical control variate 
\begin{equation}
    w_{p,1} = w_p - w_{p,0}
\end{equation}
 is  applied on the marker weights at each time step prior to the creation of the solver's right-hand side. The reduction $w_{p,0}$ for particle $p$ is  chosen to be  a Maxwellian  defined at the particle position as
\begin{equation}
    w_{p,0} = n_0(r_p)\left(\frac{2\pi T_0(r_p)}{m_p}\right)^{-1.5}\exp{\left(-\frac{1}{2}v_p^2\frac{m_s}{T_0(r_p)}\right)}
\end{equation}
with 
\begin{equation}
    v_p = \sqrt{\frac{2\mu_p B_p}{m_p}+v_{p,||}^2}
\end{equation}

\subsubsection{Results}
\begin{table}[h]
\centering
\begin{tabularx}{\textwidth}{l|XXXX}
\hline
$n_e=n_i$ [$cm^{-3}$]& $\num{4.768e13}$ & $\num{4.768e14}$ & $\num{4.768e15}$ & $\num{4.768e16}$ \\
maximum time step $\Delta t_{CFL}$ [s] & $\num{1.363e-7}$ & $\num{1.363e-7}$ & $\num{1.363e-7}$ & $\num{1.363e-7}$\\ 
$T_e=T_i$ [eV]&  $30$ & $30$ & $30$ & $30$\\
$q_0$ [-] & $2$ & $2$ & $2$ & $2$\\
Predicted frequency $\omega_{MHD}$ $[s^{-1}]$ & $\num{8.09e3}$ & $\num{2.56e4}$ & $\num{8.09e4}$ & $\num{2.56e5}$\\
Simulated frequency $\omega$ $[s^{-1}]$ for $m_i/m_e=20$ & $\num{8.007e3}$ & $\num{2.480e4}$ & $\num{6.484e4}$ & $\num{9.981e4}$\\
Simulated frequency $\omega$ $[s^{-1}]$ for $m_i/m_e=200$ & $\num{7.963e3}$ & $\num{2.539e4}$ & $\num{7.886e4}$ & $\num{2.030e5}$\\
Simulated frequency $\omega$ $[s^{-1}]$ for $m_i/m_e=2000$ & $\num{8.081e3}$ &  $\num{2.517e4}$ & $\num{7.976e4}$ & $\num{2.495e5}$\\
\hline
$n_e=n_i$ [$cm^{-3}$]& $\num{1e14}$ & $\num{1e14}$ & $\num{1e14}$ & $\num{1e14}$ \\
maximum time step $\Delta t_{CFL}$ [s] & $\num{1.363e-7}$ & $\num{1.363e-7}$ & $\num{1.363e-7}$ & $\num{1.363e-7}$ \\ 
$T_e=T_i$ [eV]&  $30$ & $30$ & $30$ & $30$\\
$q_0$ [-] & $1.1$ & $2$ & $3$ & $4$\\
Predicted frequency $\omega_{MHD}$ $[s^{-1}]$ & $\num{3.21e5}$ & $\num{1.77e5}$ & $\num{1.18e5}$ & $\num{8.84e4}$\\
Simulated frequency $\omega$ $[s^{-1}]$ for $m_i/m_e=20$ & $\num{1.680e5}$ & $\num{9.297e4}$ & $\num{6.184e4}$ & $\num{4.629e4}$\\
Simulated frequency $\omega$ $[s^{-1}]$ for $m_i/m_e=200$ &  $\num{2.855e5}$ & $\num{1.565e5}$ & $\num{1.044e5}$ & $\num{7.820e4}$\\
Simulated frequency $\omega$ $[s^{-1}]$ for $m_i/m_e=2000$ & $\num{3.155e5}$ & $\num{1.743e5}$ & $\num{1.159e5}$ & $\num{8.717e4}$\\
\hline
$n_e=n_i$ [$cm^{-3}$]& $\num{1.429e14}$ & $\num{1.429e14}$ & $\num{1.429e14}$ & $\num{1.429e14}$ \\
maximum time step $\Delta t_{CFL}$ [s] & $\num{2.371e-7}$ & $\num{7.470e-8}$ & $\num{2.371e-8}$ & $\num{7.470e-9}$ \\ 
$T_e=T_i$ [eV]&  $10$ & $100$ & $1000$ & $10000$\\
$q_0$ [-] & $2$ & $2$ & $2$ & $2$\\
Predicted frequency $\omega_{MHD}$ $[s^{-1}]$ & $\num{1.478e5}$ & $\num{1.478e5}$ & $\num{1.478e5}$ & $\num{1.478e5}$ \\
Simulated frequency $\omega$ $[s^{-1}]$ for $m_i/m_e=20$& $\num{8.709e4}$ & $\num{8.921e4}$ & $\num{1.129e5}$ & $\num{1.261e5}$\\
Simulated frequency $\omega$ $[s^{-1}]$ for $m_i/m_e=200$& $\num{1.346e5}$ & $\num{1.384e5}$ & $\num{1.511e5}$ & $\num{2.379e5}$\\
Simulated frequency $\omega$ $[s^{-1}]$ for $m_i/m_e=2000$& $\num{3.634e5}$ & $\num{1.467e5}$ & $\num{1.595e5}$ & $\num{2.522e5}$\\
Simulated growth rate $\gamma$ $[s^{-1}]$ for $m_i/m_e=20$& $\num{6.248e2}$ & $\num{-2.912e3}$ & $\num{-1.163e5}$ & $\num{-8.588e5}$\\
Simulated growth rate $\gamma$ $[s^{-1}]$ for $m_i/m_e=200$& $\num{2.064e3}$ & $\num{-1.121e4}$ & $\num{-9.542e4}$ & $\num{-3.392e5}$\\
Simulated growth rate $\gamma$ $[s^{-1}]$ for $m_i/m_e=2000$& $\num{-2.146e3}$ & $\num{-1.082e4}$ & $\num{-3.696e4}$ & $\num{-1.166e5}$\\
\hline
\end{tabularx}
\caption{Parameters $n_0$, $T_0$ and $q_0$, analytically expected frequencies according to \cref{eq:disprel_saw} and numerically obtained frequencies for SAW test cases. All tests are performed for three different mass ratios $m_i/m_e = 20$, $200$ and $2000$.}
\label{tab:para}
\end{table}

\begin{figure}
    \centering
    \includegraphics[width=\linewidth]{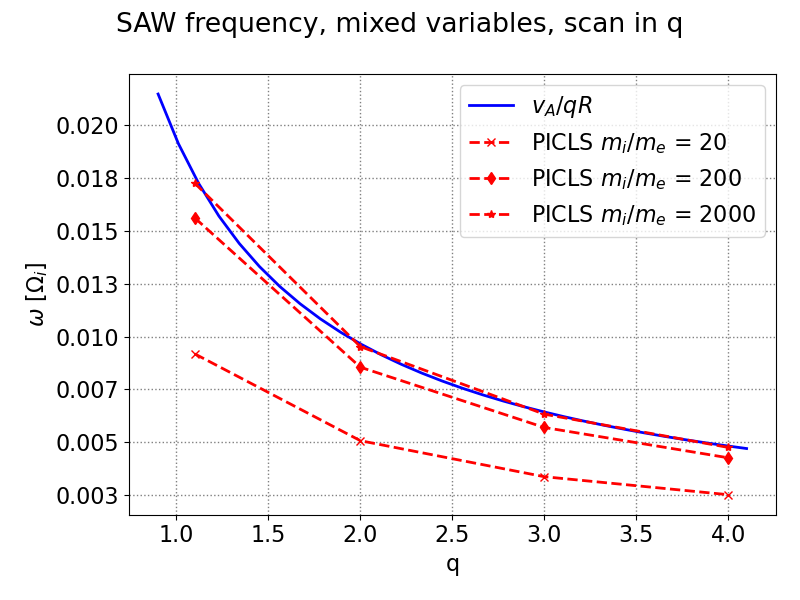}
    \caption{Shear Alfvén wave frequency observed in mixed variable PICLS for varying values of $q_0$ in units of the ion gyro frequency $\Omega_i = \num{1.83e7} Hz$.}
    \label{fig:qscan}
\end{figure}

\begin{figure}
    \centering
    \includegraphics[width=\linewidth]{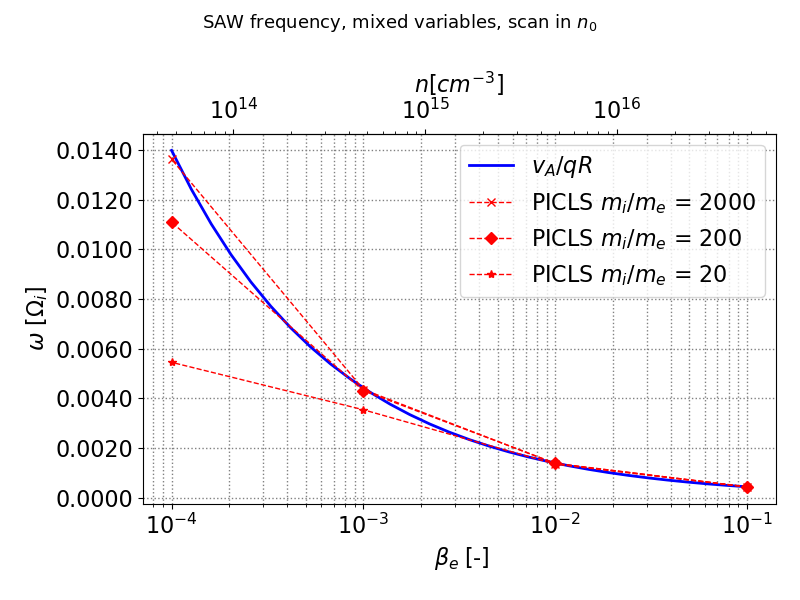}
    \caption{Shear Alfvén wave frequency observed in mixed variable PICLS for varying values of $n_0$ in units of the ion gyro frequency $\Omega_i = \num{1.83e7} Hz$.}
    \label{fig:betascan}
\end{figure}

\begin{figure}
    \centering
    \includegraphics[width=\linewidth]{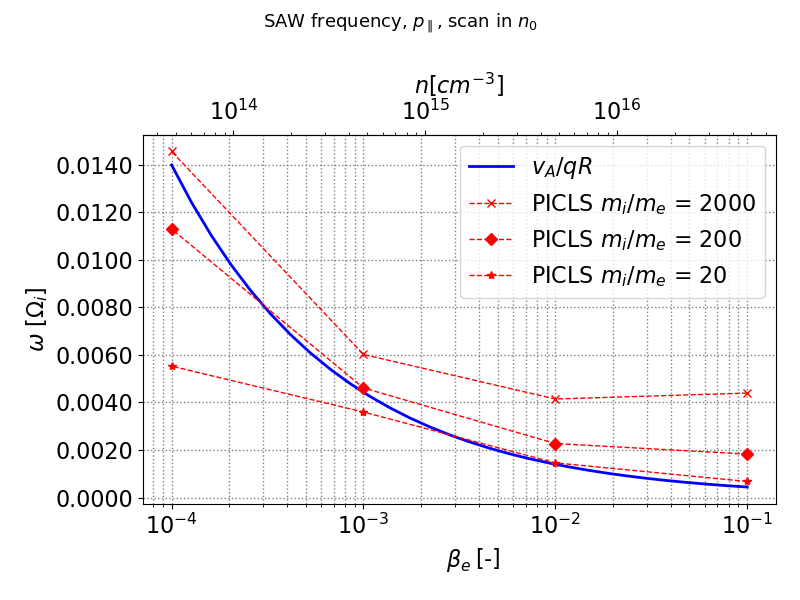}
    \caption{Shear Alfvén wave frequency observed in the $p_{\parallel}$ implementation of PICLS for varying values of $n_0$ in units of the ion gyro frequency $\Omega_i = \num{1.83e7} Hz$.}
    \label{fig:betascan_ppll}
\end{figure}

\Cref{fig:betascan} and \Cref{fig:qscan} reproduce figures 1 and 2 of ref.~\onlinecite{Biancalani2016} in very good approximation, despite a difference in the definition of $\beta$. While not present in the reference, we added data for mass ratios of 20 and 2000 to \cref{fig:betascan} to demonstrate the convergence of the low $\beta$ SAW dynamic to the analytic prediction with increasingly natural mass ratio. There is a notable deviation between the frequencies displayed in \cref{fig:betascan}, obtained using the mixed variable formulation, and the ones of \cref{fig:betascan_ppll} which result from $p_{\parallel}$-runs. The mismatch becomes most notable for high values of $\beta$ and high mass ratios. This is in agreement with the theory, which predicts increasing severity of the cancellation problem with increasing value of the skin term in \cref{eq:ampere}  if the number of used markers remains constant. Hence, the numerical error becomes largest on those data points. 


\begin{figure}
    \centering
    \includegraphics[width=\linewidth]{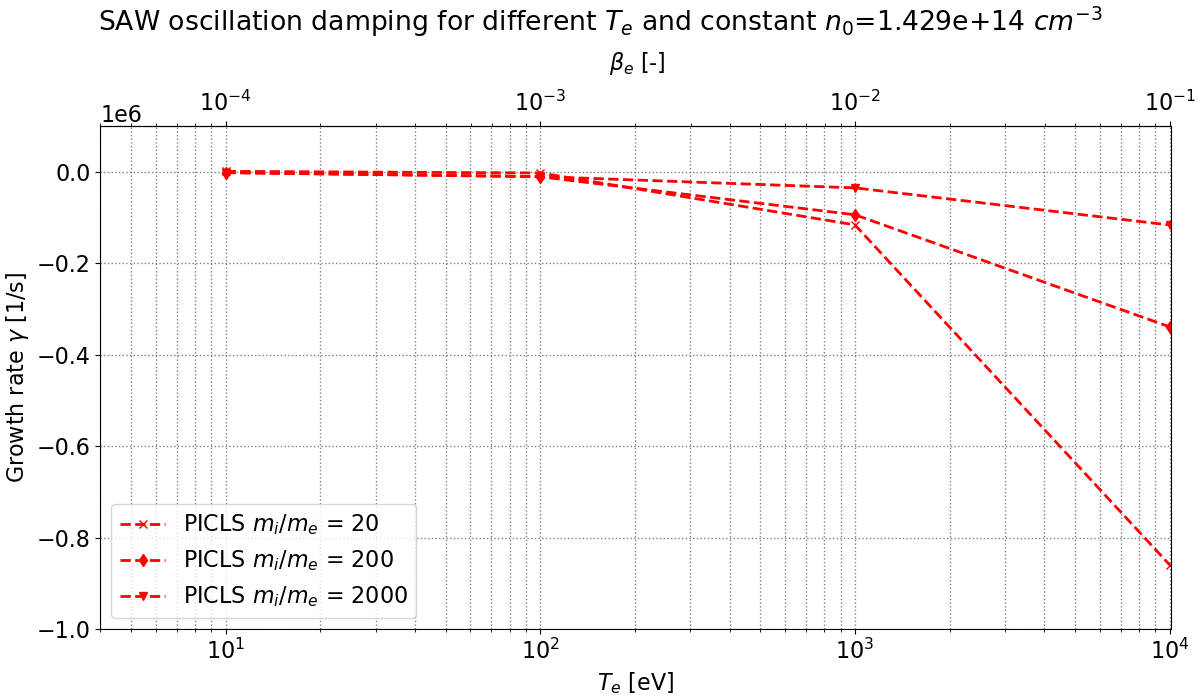}
    \caption{Growth rates of SAW oscillations observed in mixed variable PICLS for increasing $T_0$.}
    \label{fig:Tdamping}
\end{figure}

\subsection{Verification of the gyroaverage implementation using ITG simulations}
\subsubsection{Method}
To verify the gyroaveraging routine in PICLS, we expand the verification of the electrostatic version of PICLS against GENE-X results\cite{MichelsPhD} presented in ref.~\onlinecite{bottino2024} by comparing the ITG growth rates for purely driftkinetic runs with those obtained using gyrokinetic ions. Finite Larmor radius effects should influence the results the more the closer the ion Larmor radius $\rho_{L,i}$ comes to the wavelength of the ITG determined by $m$.
\subsubsection{Setup}
The geometry chosen is a cylindrical domain of radius $r_{max}=4.68$ cm and length $484$ cm in the limit of $q_0 \rightarrow \infty$. This assures that $n$ and $m$ are decoupled. The magnetic field strength is $B_0=2T$. The density profile is flat at $n_e=n_i=\num{1.5e13}$ cm$^{-3}$  while the temperature profile, which is also shown in our previous use of a similar setup for finite $q_0$\cite{bottino2024}, is given by
\begin{equation}
    T(s)=T_0\exp{\left(-\kappa_T \sigma_T \tanh{\left(\frac{s-s_0}{\sigma_T}\right)}\right)}
\end{equation}
with $s=r/r_{max}$, $\sigma_{T}=0.15$, $s_0=0.5$, $\kappa_{T}=2.8$ and $T_0 = 1000$ eV.
We use a near natural mass ratio for Deuterium with $m_i=3600$ $m_e$. Under these conditions, the initial ion Larmor radius is $\rho_{L,i}=\frac{v_{th,i}}{\Omega_i} = \frac{\num{2.210e+07} cm/s}{\num{9.771e+07} rad/s} \approx \num{0.226}$ cm. During the gyrokinetic simulation, the number of ion Larmor points is dynamically adapted to the size of $\rho_{L,i}$ within the limits of 4 and 8. After initializing a density perturbation of $n=1$ and varying $m$ to set off the mode, we let the simulation run for 200 time steps of $\Delta t=\num{8.064e-9}$ s each.

\subsubsection{Results}
\begin{figure}
    \centering
    \includegraphics[width=\linewidth]{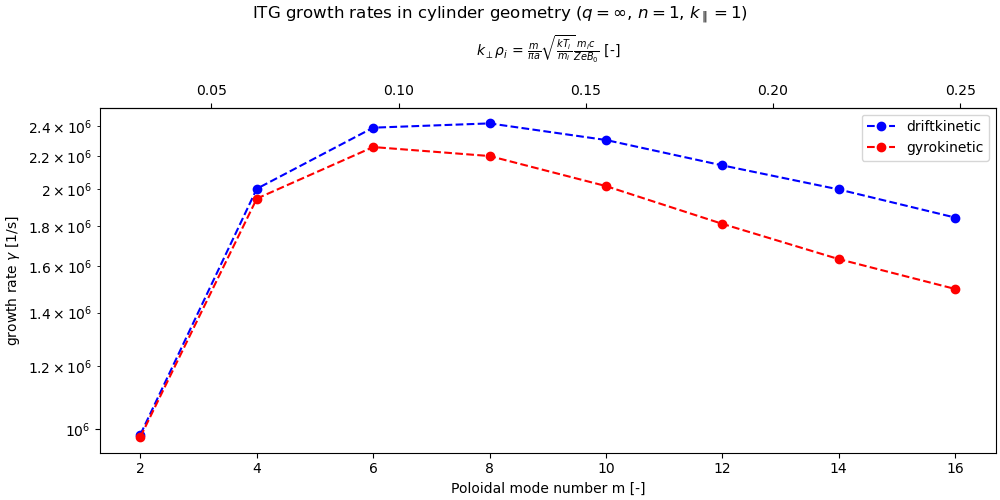}
    \caption{Growth rates of ITG oscillations observed in electrostatic PICLS with and without gyroaveraging for different poloidal mode numbers $m$.}
    \label{fig:gyrocheck}
\end{figure}
For low $m$, equalling low $k_{\perp}\rho_i$, the results for gyrokinetic and driftkinetic treatment converge as expected. Finite Larmor radius effects manifest as the difference between the gyrokinetic and driftkinetic ITG growth rates in \cref{fig:gyrocheck} as the extension of the field period in configuration space approaches the length scale of the ion Larmor radius. 

\section{Conclusion and outlook}
In an effort to include electromagnetic effects in the gyrokinetic PIC code PICLS, we added an additional field solver for the parallel magnetic vector potential $A_{||}$  and made the necessary changes to the particle pusher by using a generalized potential in the equations of motion. The newly implemented \Ampere-solver was modeled after the existing Poisson solver of the electrostatic potential\cite{Stier2024}, and the implementation was verified using the method of manufactured solutions. Since we used a $p_{\parallel}$-formulation, PICLS encounters the \Ampere-cancellation problem. To mitigate its effects, we implemented a mixed variable scheme with a pullback transformation\cite{mishchenko2014} similar to the approaches used in multiple other electromagnetic PIC codes. Using shear-Alfvén-wave simulations, we showed that the new electromagnetic version of PICLS is valid and reproduces electromagnetic physics as expected. Additionally, we demonstrated PICLS' capability to include ion finite Larmor radius effects in the example of an ITG.\\
As a next step in verifying the new electromagnetic model in PICLS, we aim to reproduce the ITG-KBM transition\cite{goerler2016,sturdevant2021} performed by various other electromagnetic codes in the past as a benchmark. This will require the first use of a circular torus geometry in PICLS. Apart from that, we plan to implement an Ohm's law closure\cite{mishchenko2014_2} to the mixed variable scheme as a third way to run electromagnetic simulations. While the currently implemented version of the mixed variable approach allows for electromagnetic simulations at more reasonable marker counts than the unmitigated $p_{\parallel}$-formulation, an Ohm closure additionally alleviates the time step constraints.   

\section{Acknowledgement}
This work has been carried out within the framework of the EUROfusion Consortium, funded by the European Union via the Euratom Research and Training Programme (Grant Agreement No 101052200 — EUROfusion). Views and opinions expressed are however those of the author(s) only and do not necessarily reflect those of the European Union or the European Commission. Neither the European Union nor the European Commission can be held responsible for them.\\  The Swiss contribution to this work has been funded by the Swiss State Secretariat for Education, Research and Innovation (SERI). Views and opinions expressed are however those of the author(s) only and do not necessarily reflect those of the European Union, the European Commission or SERI. Neither the European Union nor the European Commission nor SERI can be held responsible for them.

\bibliography{bib2.bib}

\begin{thebibliography}{25}%
\makeatletter
\providecommand \@ifxundefined [1]{%
 \@ifx{#1\undefined}
}%
\providecommand \@ifnum [1]{%
 \ifnum #1\expandafter \@firstoftwo
 \else \expandafter \@secondoftwo
 \fi
}%
\providecommand \@ifx [1]{%
 \ifx #1\expandafter \@firstoftwo
 \else \expandafter \@secondoftwo
 \fi
}%
\providecommand \natexlab [1]{#1}%
\providecommand \enquote  [1]{``#1''}%
\providecommand \bibnamefont  [1]{#1}%
\providecommand \bibfnamefont [1]{#1}%
\providecommand \citenamefont [1]{#1}%
\providecommand \href@noop [0]{\@secondoftwo}%
\providecommand \href [0]{\begingroup \@sanitize@url \@href}%
\providecommand \@href[1]{\@@startlink{#1}\@@href}%
\providecommand \@@href[1]{\endgroup#1\@@endlink}%
\providecommand \@sanitize@url [0]{\catcode `\\12\catcode `\$12\catcode
  `\&12\catcode `\#12\catcode `\^12\catcode `\_12\catcode `\%12\relax}%
\providecommand \@@startlink[1]{}%
\providecommand \@@endlink[0]{}%
\providecommand \url  [0]{\begingroup\@sanitize@url \@url }%
\providecommand \@url [1]{\endgroup\@href {#1}{\urlprefix }}%
\providecommand \urlprefix  [0]{URL }%
\providecommand \Eprint [0]{\href }%
\providecommand \doibase [0]{http://dx.doi.org/}%
\providecommand \selectlanguage [0]{\@gobble}%
\providecommand \bibinfo  [0]{\@secondoftwo}%
\providecommand \bibfield  [0]{\@secondoftwo}%
\providecommand \translation [1]{[#1]}%
\providecommand \BibitemOpen [0]{}%
\providecommand \bibitemStop [0]{}%
\providecommand \bibitemNoStop [0]{.\EOS\space}%
\providecommand \EOS [0]{\spacefactor3000\relax}%
\providecommand \BibitemShut  [1]{\csname bibitem#1\endcsname}%
\let\auto@bib@innerbib\@empty
\bibitem [{\citenamefont {Boesl}(2020)}]{MatthiasPhD}%
  \BibitemOpen
  \bibfield  {author} {\bibinfo {author} {\bibfnamefont {M.}~\bibnamefont
  {Boesl}},\ }\emph {\bibinfo {title} {{PICLS: a gyrokinetic full-f
  particle-in-cell code for the scrape-off layer}}},\ \href@noop {} {Ph.D.
  thesis},\ \bibinfo  {school} {Technical University of Munich} (\bibinfo
  {year} {2020})\BibitemShut {NoStop}%
\bibitem [{\citenamefont {Bottino}\ \emph {et~al.}(2024)\citenamefont
  {Bottino}, \citenamefont {Stier}, \citenamefont {Boesl}, \citenamefont
  {Hayward-Schneider}, \citenamefont {Bergmann}, \citenamefont {Coster},
  \citenamefont {Brunner},\ and\ \citenamefont {Villard}}]{bottino2024}%
  \BibitemOpen
  \bibfield  {author} {\bibinfo {author} {\bibfnamefont {A.}~\bibnamefont
  {Bottino}}, \bibinfo {author} {\bibfnamefont {A.}~\bibnamefont {Stier}},
  \bibinfo {author} {\bibfnamefont {M.}~\bibnamefont {Boesl}}, \bibinfo
  {author} {\bibfnamefont {T.}~\bibnamefont {Hayward-Schneider}}, \bibinfo
  {author} {\bibfnamefont {A.}~\bibnamefont {Bergmann}}, \bibinfo {author}
  {\bibfnamefont {D.}~\bibnamefont {Coster}}, \bibinfo {author} {\bibfnamefont
  {S.}~\bibnamefont {Brunner}}, \ and\ \bibinfo {author} {\bibfnamefont
  {L.}~\bibnamefont {Villard}},\ }\href@noop {} {\bibfield  {journal} {\bibinfo
   {journal} {Plasma Physics and Controlled Fusion}\ } (\bibinfo {year}
  {2024})}\BibitemShut {NoStop}%
\bibitem [{\citenamefont {Stier}\ \emph {et~al.}(2024)\citenamefont {Stier},
  \citenamefont {Bottino}, \citenamefont {Boesl}, \citenamefont {Pinto},
  \citenamefont {Hayward-Schneider}, \citenamefont {Coster}, \citenamefont
  {Bergmann}, \citenamefont {Murugappan}, \citenamefont {Brunner},
  \citenamefont {Villard},\ and\ \citenamefont {Jenko}}]{Stier2024}%
  \BibitemOpen
  \bibfield  {author} {\bibinfo {author} {\bibfnamefont {A.}~\bibnamefont
  {Stier}}, \bibinfo {author} {\bibfnamefont {A.}~\bibnamefont {Bottino}},
  \bibinfo {author} {\bibfnamefont {M.}~\bibnamefont {Boesl}}, \bibinfo
  {author} {\bibfnamefont {M.~C.}\ \bibnamefont {Pinto}}, \bibinfo {author}
  {\bibfnamefont {T.}~\bibnamefont {Hayward-Schneider}}, \bibinfo {author}
  {\bibfnamefont {D.}~\bibnamefont {Coster}}, \bibinfo {author} {\bibfnamefont
  {A.}~\bibnamefont {Bergmann}}, \bibinfo {author} {\bibfnamefont
  {M.}~\bibnamefont {Murugappan}}, \bibinfo {author} {\bibfnamefont
  {S.}~\bibnamefont {Brunner}}, \bibinfo {author} {\bibfnamefont
  {L.}~\bibnamefont {Villard}}, \ and\ \bibinfo {author} {\bibfnamefont
  {F.}~\bibnamefont {Jenko}},\ }\href@noop {} {\bibfield  {journal} {\bibinfo
  {journal} {Computer Physics Communications}\ }\textbf {\bibinfo {volume}
  {299}} (\bibinfo {year} {2024})}\BibitemShut {NoStop}%
\bibitem [{\citenamefont {Hatzky}(2006)}]{Hatzky06}%
  \BibitemOpen
  \bibfield  {author} {\bibinfo {author} {\bibfnamefont {R.}~\bibnamefont
  {Hatzky}},\ }\href@noop {} {\bibfield  {journal} {\bibinfo  {journal}
  {Parallel Computing}\ }\textbf {\bibinfo {volume} {32}},\ \bibinfo {pages}
  {325} (\bibinfo {year} {2006})}\BibitemShut {NoStop}%
\bibitem [{\citenamefont {Lanti}\ \emph {et~al.}(2020)\citenamefont {Lanti},
  \citenamefont {Ohana}, \citenamefont {Tronko}, \citenamefont
  {Hayward-Schneider}, \citenamefont {Bottino}, \citenamefont {McMillan},
  \citenamefont {Mishchenko}, \citenamefont {Scheinberg}, \citenamefont
  {Biancalani}, \citenamefont {Angelino}, \citenamefont {Brunner},
  \citenamefont {Dominski}, \citenamefont {Donnel}, \citenamefont {Gheller},
  \citenamefont {Hatzky}, \citenamefont {Jocksch}, \citenamefont {Jolliet},
  \citenamefont {Lu}, \citenamefont {Collar}, \citenamefont {Novikau},
  \citenamefont {Sonnendruecker}, \citenamefont {Vernay},\ and\ \citenamefont
  {Villard}}]{LantiCPC2020}%
  \BibitemOpen
  \bibfield  {author} {\bibinfo {author} {\bibfnamefont {E.}~\bibnamefont
  {Lanti}}, \bibinfo {author} {\bibfnamefont {N.}~\bibnamefont {Ohana}},
  \bibinfo {author} {\bibfnamefont {N.}~\bibnamefont {Tronko}}, \bibinfo
  {author} {\bibfnamefont {T.}~\bibnamefont {Hayward-Schneider}}, \bibinfo
  {author} {\bibfnamefont {A.}~\bibnamefont {Bottino}}, \bibinfo {author}
  {\bibfnamefont {B.~F.}\ \bibnamefont {McMillan}}, \bibinfo {author}
  {\bibfnamefont {A.}~\bibnamefont {Mishchenko}}, \bibinfo {author}
  {\bibfnamefont {A.}~\bibnamefont {Scheinberg}}, \bibinfo {author}
  {\bibfnamefont {A.}~\bibnamefont {Biancalani}}, \bibinfo {author}
  {\bibfnamefont {P.}~\bibnamefont {Angelino}}, \bibinfo {author}
  {\bibfnamefont {S.}~\bibnamefont {Brunner}}, \bibinfo {author} {\bibfnamefont
  {J.}~\bibnamefont {Dominski}}, \bibinfo {author} {\bibfnamefont
  {P.}~\bibnamefont {Donnel}}, \bibinfo {author} {\bibfnamefont
  {C.}~\bibnamefont {Gheller}}, \bibinfo {author} {\bibfnamefont
  {R.}~\bibnamefont {Hatzky}}, \bibinfo {author} {\bibfnamefont
  {A.}~\bibnamefont {Jocksch}}, \bibinfo {author} {\bibfnamefont
  {S.}~\bibnamefont {Jolliet}}, \bibinfo {author} {\bibfnamefont {Z.~X.}\
  \bibnamefont {Lu}}, \bibinfo {author} {\bibfnamefont {J.~P.~M.}\ \bibnamefont
  {Collar}}, \bibinfo {author} {\bibfnamefont {I.}~\bibnamefont {Novikau}},
  \bibinfo {author} {\bibfnamefont {E.}~\bibnamefont {Sonnendruecker}},
  \bibinfo {author} {\bibfnamefont {T.}~\bibnamefont {Vernay}}, \ and\ \bibinfo
  {author} {\bibfnamefont {L.}~\bibnamefont {Villard}},\ }\href {\doibase
  {10.1016/j.cpc.2019.107072}} {\bibfield  {journal} {\bibinfo  {journal}
  {Computer Physics Communications}\ }\textbf {\bibinfo {volume} {{251}}}
  (\bibinfo {year} {2020}),\ {10.1016/j.cpc.2019.107072}}\BibitemShut {NoStop}%
\bibitem [{\citenamefont {Boesl}\ \emph {et~al.}(2019)\citenamefont {Boesl},
  \citenamefont {Bergmann}, \citenamefont {Bottino}, \citenamefont {Coster},
  \citenamefont {Lanti}, \citenamefont {Ohana},\ and\ \citenamefont
  {Jenko}}]{Boesl2019}%
  \BibitemOpen
  \bibfield  {author} {\bibinfo {author} {\bibfnamefont {M.}~\bibnamefont
  {Boesl}}, \bibinfo {author} {\bibfnamefont {A.}~\bibnamefont {Bergmann}},
  \bibinfo {author} {\bibfnamefont {A.}~\bibnamefont {Bottino}}, \bibinfo
  {author} {\bibfnamefont {D.}~\bibnamefont {Coster}}, \bibinfo {author}
  {\bibfnamefont {E.}~\bibnamefont {Lanti}}, \bibinfo {author} {\bibfnamefont
  {N.}~\bibnamefont {Ohana}}, \ and\ \bibinfo {author} {\bibfnamefont
  {F.}~\bibnamefont {Jenko}},\ }\href {\doibase 10.1063/1.5121262} {\bibfield
  {journal} {\bibinfo  {journal} {Physics of Plasmas}\ }\textbf {\bibinfo
  {volume} {26}},\ \bibinfo {pages} {122302} (\bibinfo {year}
  {2019})}\BibitemShut {NoStop}%
\bibitem [{\citenamefont {Mishchenko}\ \emph
  {et~al.}(2014{\natexlab{a}})\citenamefont {Mishchenko}, \citenamefont
  {Könies}, \citenamefont {Kleiber},\ and\ \citenamefont
  {Cole}}]{mishchenko2014}%
  \BibitemOpen
  \bibfield  {author} {\bibinfo {author} {\bibfnamefont {A.}~\bibnamefont
  {Mishchenko}}, \bibinfo {author} {\bibfnamefont {A.}~\bibnamefont {Könies}},
  \bibinfo {author} {\bibfnamefont {R.}~\bibnamefont {Kleiber}}, \ and\
  \bibinfo {author} {\bibfnamefont {M.}~\bibnamefont {Cole}},\ }\href@noop {}
  {\bibfield  {journal} {\bibinfo  {journal} {Physics of Plasmas}\ }\textbf
  {\bibinfo {volume} {21}} (\bibinfo {year} {2014}{\natexlab{a}})}\BibitemShut
  {NoStop}%
\bibitem [{\citenamefont {Cole}(2016)}]{cole2016}%
  \BibitemOpen
  \bibfield  {author} {\bibinfo {author} {\bibfnamefont {M.~D.~J.}\
  \bibnamefont {Cole}},\ }\emph {\bibinfo {title} {{Global gyrokinetic and
  fluid hybrid simulations of tokamaks and stellarators}}},\ \href@noop {}
  {Ph.D. thesis},\ \bibinfo  {school} {Ernst-Moritz-Arndt-University
  Greifswald} (\bibinfo {year} {2016})\BibitemShut {NoStop}%
\bibitem [{\citenamefont {Mishchenko}\ \emph
  {et~al.}(2019{\natexlab{a}})\citenamefont {Mishchenko}, \citenamefont
  {Bottino}, \citenamefont {Biancalani}, \citenamefont {Hatzky}, \citenamefont
  {Hayward-Schneider}, \citenamefont {Ohana}, \citenamefont {Lanti},
  \citenamefont {Brunner}, \citenamefont {Villard}, \citenamefont {Borchardt},
  \citenamefont {Kleiber},\ and\ \citenamefont {Könies}}]{mishchenko2019}%
  \BibitemOpen
  \bibfield  {author} {\bibinfo {author} {\bibfnamefont {A.}~\bibnamefont
  {Mishchenko}}, \bibinfo {author} {\bibfnamefont {A.}~\bibnamefont {Bottino}},
  \bibinfo {author} {\bibfnamefont {A.}~\bibnamefont {Biancalani}}, \bibinfo
  {author} {\bibfnamefont {R.}~\bibnamefont {Hatzky}}, \bibinfo {author}
  {\bibfnamefont {T.}~\bibnamefont {Hayward-Schneider}}, \bibinfo {author}
  {\bibfnamefont {N.}~\bibnamefont {Ohana}}, \bibinfo {author} {\bibfnamefont
  {E.}~\bibnamefont {Lanti}}, \bibinfo {author} {\bibfnamefont
  {S.}~\bibnamefont {Brunner}}, \bibinfo {author} {\bibfnamefont
  {L.}~\bibnamefont {Villard}}, \bibinfo {author} {\bibfnamefont
  {M.}~\bibnamefont {Borchardt}}, \bibinfo {author} {\bibfnamefont
  {R.}~\bibnamefont {Kleiber}}, \ and\ \bibinfo {author} {\bibfnamefont
  {A.}~\bibnamefont {Könies}},\ }\href@noop {} {\bibfield  {journal} {\bibinfo
   {journal} {Computer Physics Communications}\ }\textbf {\bibinfo {volume}
  {238}},\ \bibinfo {pages} {194} (\bibinfo {year}
  {2019}{\natexlab{a}})}\BibitemShut {NoStop}%
\bibitem [{\citenamefont {Hager}\ \emph {et~al.}(2022)\citenamefont {Hager},
  \citenamefont {Ku}, \citenamefont {Sharma}, \citenamefont {Chang},\ and\
  \citenamefont {Churchill}}]{hager2022}%
  \BibitemOpen
  \bibfield  {author} {\bibinfo {author} {\bibfnamefont {R.}~\bibnamefont
  {Hager}}, \bibinfo {author} {\bibfnamefont {S.-H.}\ \bibnamefont {Ku}},
  \bibinfo {author} {\bibfnamefont {A.~Y.}\ \bibnamefont {Sharma}}, \bibinfo
  {author} {\bibfnamefont {C.~S.}\ \bibnamefont {Chang}}, \ and\ \bibinfo
  {author} {\bibfnamefont {R.~M.}\ \bibnamefont {Churchill}},\ }\href@noop {}
  {\bibfield  {journal} {\bibinfo  {journal} {Physics of Plasmas}\ }\textbf
  {\bibinfo {volume} {29}} (\bibinfo {year} {2022})}\BibitemShut {NoStop}%
\bibitem [{\citenamefont {Biancalani}\ \emph {et~al.}(2016)\citenamefont
  {Biancalani}, \citenamefont {Bottino}, \citenamefont {Briguglio},
  \citenamefont {Könies}, \citenamefont {Lauber}, \citenamefont {Mishchenko},
  \citenamefont {Poli}, \citenamefont {Scott},\ and\ \citenamefont
  {Zonca}}]{Biancalani2016}%
  \BibitemOpen
  \bibfield  {author} {\bibinfo {author} {\bibfnamefont {A.}~\bibnamefont
  {Biancalani}}, \bibinfo {author} {\bibfnamefont {A.}~\bibnamefont {Bottino}},
  \bibinfo {author} {\bibfnamefont {S.}~\bibnamefont {Briguglio}}, \bibinfo
  {author} {\bibfnamefont {A.}~\bibnamefont {Könies}}, \bibinfo {author}
  {\bibfnamefont {P.}~\bibnamefont {Lauber}}, \bibinfo {author} {\bibfnamefont
  {A.}~\bibnamefont {Mishchenko}}, \bibinfo {author} {\bibfnamefont
  {E.}~\bibnamefont {Poli}}, \bibinfo {author} {\bibfnamefont {B.~D.}\
  \bibnamefont {Scott}}, \ and\ \bibinfo {author} {\bibfnamefont
  {F.}~\bibnamefont {Zonca}},\ }\href {\doibase 10.1063/1.4939803} {\bibfield
  {journal} {\bibinfo  {journal} {Physics of Plasmas}\ }\textbf {\bibinfo
  {volume} {23}},\ \bibinfo {pages} {012108} (\bibinfo {year}
  {2016})}\BibitemShut {NoStop}%
\bibitem [{\citenamefont {Scott}(2021)}]{ScottV2}%
  \BibitemOpen
  \bibfield  {author} {\bibinfo {author} {\bibfnamefont {B.}~\bibnamefont
  {Scott}},\ }\href {\doibase 10.1088/978-0-7503-3855-4} {\emph {\bibinfo
  {title} {Turbulence and Instabilities in Magnetised Plasmas, Volume 2}}},\
  2053-2563\ (\bibinfo  {publisher} {IOP Publishing},\ \bibinfo {year}
  {2021})\BibitemShut {NoStop}%
\bibitem [{\citenamefont {Tronko}, \citenamefont {Bottino},\ and\ \citenamefont
  {Sonnendr{\"u}cker}(2016)}]{Tronko16}%
  \BibitemOpen
  \bibfield  {author} {\bibinfo {author} {\bibfnamefont {N.}~\bibnamefont
  {Tronko}}, \bibinfo {author} {\bibfnamefont {A.}~\bibnamefont {Bottino}}, \
  and\ \bibinfo {author} {\bibfnamefont {E.}~\bibnamefont
  {Sonnendr{\"u}cker}},\ }\href@noop {} {\bibfield  {journal} {\bibinfo
  {journal} {Physics of Plasmas}\ }\textbf {\bibinfo {volume} {23}},\ \bibinfo
  {pages} {082505} (\bibinfo {year} {2016})}\BibitemShut {NoStop}%
\bibitem [{\citenamefont {Bottino}\ and\ \citenamefont
  {Sonnendr{\"u}cker}(2015)}]{Bott15}%
  \BibitemOpen
  \bibfield  {author} {\bibinfo {author} {\bibfnamefont {A.}~\bibnamefont
  {Bottino}}\ and\ \bibinfo {author} {\bibfnamefont {E.}~\bibnamefont
  {Sonnendr{\"u}cker}},\ }\href@noop {} {\bibfield  {journal} {\bibinfo
  {journal} {Journal of Plasma Physics}\ }\textbf {\bibinfo {volume} {81}}
  (\bibinfo {year} {2015})}\BibitemShut {NoStop}%
\bibitem [{\citenamefont {Tronko}\ and\ \citenamefont
  {Chandre}(2018)}]{Tronko18}%
  \BibitemOpen
  \bibfield  {author} {\bibinfo {author} {\bibfnamefont {N.}~\bibnamefont
  {Tronko}}\ and\ \bibinfo {author} {\bibfnamefont {C.}~\bibnamefont
  {Chandre}},\ }\href@noop {} {\bibfield  {journal} {\bibinfo  {journal}
  {Journal of Plasma Physics}\ }\textbf {\bibinfo {volume} {84}} (\bibinfo
  {year} {2018})}\BibitemShut {NoStop}%
\bibitem [{\citenamefont {Mishchenko}\ \emph
  {et~al.}(2014{\natexlab{b}})\citenamefont {Mishchenko}, \citenamefont {Cole},
  \citenamefont {Kleiber},\ and\ \citenamefont {Könies}}]{mishchenko2014_2}%
  \BibitemOpen
  \bibfield  {author} {\bibinfo {author} {\bibfnamefont {A.}~\bibnamefont
  {Mishchenko}}, \bibinfo {author} {\bibfnamefont {M.}~\bibnamefont {Cole}},
  \bibinfo {author} {\bibfnamefont {R.}~\bibnamefont {Kleiber}}, \ and\
  \bibinfo {author} {\bibfnamefont {A.}~\bibnamefont {Könies}},\ }\href@noop
  {} {\bibfield  {journal} {\bibinfo  {journal} {Physics of Plasmas}\ }\textbf
  {\bibinfo {volume} {21}} (\bibinfo {year} {2014}{\natexlab{b}})}\BibitemShut
  {NoStop}%
\bibitem [{\citenamefont {Mishchenko}\ \emph
  {et~al.}(2019{\natexlab{b}})\citenamefont {Mishchenko}, \citenamefont
  {Bottino}, \citenamefont {Biancalani}, \citenamefont {Hatzky}, \citenamefont
  {Hayward-Schneider}, \citenamefont {Ohana}, \citenamefont {Lanti},
  \citenamefont {Brunner}, \citenamefont {Villard}, \citenamefont {Borchardt}
  \emph {et~al.}}]{mishchenko19}%
  \BibitemOpen
  \bibfield  {author} {\bibinfo {author} {\bibfnamefont {A.}~\bibnamefont
  {Mishchenko}}, \bibinfo {author} {\bibfnamefont {A.}~\bibnamefont {Bottino}},
  \bibinfo {author} {\bibfnamefont {A.}~\bibnamefont {Biancalani}}, \bibinfo
  {author} {\bibfnamefont {R.}~\bibnamefont {Hatzky}}, \bibinfo {author}
  {\bibfnamefont {T.}~\bibnamefont {Hayward-Schneider}}, \bibinfo {author}
  {\bibfnamefont {N.}~\bibnamefont {Ohana}}, \bibinfo {author} {\bibfnamefont
  {E.}~\bibnamefont {Lanti}}, \bibinfo {author} {\bibfnamefont
  {S.}~\bibnamefont {Brunner}}, \bibinfo {author} {\bibfnamefont
  {L.}~\bibnamefont {Villard}}, \bibinfo {author} {\bibfnamefont
  {M.}~\bibnamefont {Borchardt}},  \emph {et~al.},\ }\href@noop {} {\bibfield
  {journal} {\bibinfo  {journal} {Computer Physics Communications}\ }\textbf
  {\bibinfo {volume} {238}},\ \bibinfo {pages} {194} (\bibinfo {year}
  {2019}{\natexlab{b}})}\BibitemShut {NoStop}%
\bibitem [{\citenamefont {Bottino}\ \emph {et~al.}(2022)\citenamefont
  {Bottino}, \citenamefont {Falessi}, \citenamefont {Hayward-Schneider},
  \citenamefont {Biancalani}, \citenamefont {Briguglio}, \citenamefont
  {Hatzky}, \citenamefont {Lauber}, \citenamefont {Mishchenko}, \citenamefont
  {Poli}, \citenamefont {Rettino}, \citenamefont {Vannini}, \citenamefont
  {Wang},\ and\ \citenamefont {Zonca}}]{Bottino2022}%
  \BibitemOpen
  \bibfield  {author} {\bibinfo {author} {\bibfnamefont {A.}~\bibnamefont
  {Bottino}}, \bibinfo {author} {\bibfnamefont {M.}~\bibnamefont {Falessi}},
  \bibinfo {author} {\bibfnamefont {T.}~\bibnamefont {Hayward-Schneider}},
  \bibinfo {author} {\bibfnamefont {A.}~\bibnamefont {Biancalani}}, \bibinfo
  {author} {\bibfnamefont {S.}~\bibnamefont {Briguglio}}, \bibinfo {author}
  {\bibfnamefont {R.}~\bibnamefont {Hatzky}}, \bibinfo {author} {\bibfnamefont
  {P.}~\bibnamefont {Lauber}}, \bibinfo {author} {\bibfnamefont
  {A.}~\bibnamefont {Mishchenko}}, \bibinfo {author} {\bibfnamefont
  {E.}~\bibnamefont {Poli}}, \bibinfo {author} {\bibfnamefont {B.}~\bibnamefont
  {Rettino}}, \bibinfo {author} {\bibfnamefont {F.}~\bibnamefont {Vannini}},
  \bibinfo {author} {\bibfnamefont {X.}~\bibnamefont {Wang}}, \ and\ \bibinfo
  {author} {\bibfnamefont {F.}~\bibnamefont {Zonca}},\ }\href {\doibase
  10.1088/1742-6596/2397/1/012019} {\bibfield  {journal} {\bibinfo  {journal}
  {Journal of Physics: Conference Series}\ }\textbf {\bibinfo {volume}
  {2397}},\ \bibinfo {pages} {012019} (\bibinfo {year} {2022})}\BibitemShut
  {NoStop}%
\bibitem [{\citenamefont {Roache}(2001)}]{Roache2001}%
  \BibitemOpen
  \bibfield  {author} {\bibinfo {author} {\bibfnamefont {P.~J.}\ \bibnamefont
  {Roache}},\ }\href@noop {} {\bibfield  {journal} {\bibinfo  {journal}
  {Journal of Fluids Engineering}\ }\textbf {\bibinfo {volume} {124}},\
  \bibinfo {pages} {4} (\bibinfo {year} {2001})}\BibitemShut {NoStop}%
\bibitem [{\citenamefont {Oberkampf}\ and\ \citenamefont
  {Roy}(2010)}]{Oberkampf10}%
  \BibitemOpen
  \bibfield  {author} {\bibinfo {author} {\bibfnamefont {W.~L.}\ \bibnamefont
  {Oberkampf}}\ and\ \bibinfo {author} {\bibfnamefont {C.~J.}\ \bibnamefont
  {Roy}},\ }\href@noop {} {\emph {\bibinfo {title} {Verification and validation
  in scientific computing}}}\ (\bibinfo  {publisher} {Cambridge University
  Press},\ \bibinfo {year} {2010})\BibitemShut {NoStop}%
\bibitem [{\citenamefont {Li}\ \emph {et~al.}(2010)\citenamefont {Li},
  \citenamefont {Melenk}, \citenamefont {Wohlmuth},\ and\ \citenamefont
  {Zou}}]{Li2010}%
  \BibitemOpen
  \bibfield  {author} {\bibinfo {author} {\bibfnamefont {J.}~\bibnamefont
  {Li}}, \bibinfo {author} {\bibfnamefont {J.~M.}\ \bibnamefont {Melenk}},
  \bibinfo {author} {\bibfnamefont {B.}~\bibnamefont {Wohlmuth}}, \ and\
  \bibinfo {author} {\bibfnamefont {J.}~\bibnamefont {Zou}},\ }\href@noop {}
  {\bibfield  {journal} {\bibinfo  {journal} {Applied Numerical Mathematics}\
  }\textbf {\bibinfo {volume} {60}},\ \bibinfo {pages} {19} (\bibinfo {year}
  {2010})}\BibitemShut {NoStop}%
\bibitem [{\citenamefont {Villard}, \citenamefont {Brunner},\ and\
  \citenamefont {Vaclavik}(1995)}]{villard1995}%
  \BibitemOpen
  \bibfield  {author} {\bibinfo {author} {\bibfnamefont {L.}~\bibnamefont
  {Villard}}, \bibinfo {author} {\bibfnamefont {S.}~\bibnamefont {Brunner}}, \
  and\ \bibinfo {author} {\bibfnamefont {J.}~\bibnamefont {Vaclavik}},\
  }\href@noop {} {\bibfield  {journal} {\bibinfo  {journal} {Nuclear Fusion}\
  }\textbf {\bibinfo {volume} {35}} (\bibinfo {year} {1995})}\BibitemShut
  {NoStop}%
\bibitem [{\citenamefont {Michels}(2021)}]{MichelsPhD}%
  \BibitemOpen
  \bibfield  {author} {\bibinfo {author} {\bibfnamefont {D.}~\bibnamefont
  {Michels}},\ }\emph {\bibinfo {title} {Development of a high-peformance
  gyrokinetic turbulence code for the edge and scrape-off layer of magnetic
  confinement fusion devices}},\ \href@noop {} {Ph.D. thesis},\ \bibinfo
  {school} {"Technical University of Munich"} (\bibinfo {year}
  {2021})\BibitemShut {NoStop}%
\bibitem [{\citenamefont {Görler}\ \emph {et~al.}(2016)\citenamefont
  {Görler}, \citenamefont {Tronko}, \citenamefont {Hornsby}, \citenamefont
  {Bottino}, \citenamefont {Kleiber}, \citenamefont {Norscini}, \citenamefont
  {Grandgirard}, \citenamefont {Jenko},\ and\ \citenamefont
  {Sonnendrücker}}]{goerler2016}%
  \BibitemOpen
  \bibfield  {author} {\bibinfo {author} {\bibfnamefont {T.}~\bibnamefont
  {Görler}}, \bibinfo {author} {\bibfnamefont {N.}~\bibnamefont {Tronko}},
  \bibinfo {author} {\bibfnamefont {W.~A.}\ \bibnamefont {Hornsby}}, \bibinfo
  {author} {\bibfnamefont {A.}~\bibnamefont {Bottino}}, \bibinfo {author}
  {\bibfnamefont {R.}~\bibnamefont {Kleiber}}, \bibinfo {author} {\bibfnamefont
  {C.}~\bibnamefont {Norscini}}, \bibinfo {author} {\bibfnamefont
  {V.}~\bibnamefont {Grandgirard}}, \bibinfo {author} {\bibfnamefont
  {F.}~\bibnamefont {Jenko}}, \ and\ \bibinfo {author} {\bibfnamefont
  {E.}~\bibnamefont {Sonnendrücker}},\ }\href@noop {} {\bibfield  {journal}
  {\bibinfo  {journal} {Physics of Plasmas}\ }\textbf {\bibinfo {volume} {23}}
  (\bibinfo {year} {2016})}\BibitemShut {NoStop}%
\bibitem [{\citenamefont {Sturdevant}\ \emph {et~al.}(2021)\citenamefont
  {Sturdevant}, \citenamefont {Ku}, \citenamefont {Chacón}, \citenamefont
  {Chen}, \citenamefont {Hatch}, \citenamefont {Cole}, \citenamefont {Sharma},
  \citenamefont {Adams}, \citenamefont {Chang}, \citenamefont {Parker},\ and\
  \citenamefont {Hager}}]{sturdevant2021}%
  \BibitemOpen
  \bibfield  {author} {\bibinfo {author} {\bibfnamefont {B.~J.}\ \bibnamefont
  {Sturdevant}}, \bibinfo {author} {\bibfnamefont {S.}~\bibnamefont {Ku}},
  \bibinfo {author} {\bibfnamefont {L.}~\bibnamefont {Chacón}}, \bibinfo
  {author} {\bibfnamefont {Y.}~\bibnamefont {Chen}}, \bibinfo {author}
  {\bibfnamefont {D.}~\bibnamefont {Hatch}}, \bibinfo {author} {\bibfnamefont
  {M.~D.~J.}\ \bibnamefont {Cole}}, \bibinfo {author} {\bibfnamefont {A.~Y.}\
  \bibnamefont {Sharma}}, \bibinfo {author} {\bibfnamefont {M.~F.}\
  \bibnamefont {Adams}}, \bibinfo {author} {\bibfnamefont {C.~S.}\ \bibnamefont
  {Chang}}, \bibinfo {author} {\bibfnamefont {S.~E.}\ \bibnamefont {Parker}}, \
  and\ \bibinfo {author} {\bibfnamefont {R.}~\bibnamefont {Hager}},\
  }\href@noop {} {\bibfield  {journal} {\bibinfo  {journal} {Physics of
  Plasmas}\ }\textbf {\bibinfo {volume} {28}} (\bibinfo {year}
  {2021})}\BibitemShut {NoStop}%
\end{thebibliography}%
\end{document}